\documentclass[usenatbibt]{aa}  

\usepackage{graphicx}
\usepackage{color}
\usepackage{txfonts}
\usepackage[]{hyperref}
\usepackage{lscape}
\usepackage{multirow}
\usepackage{subcaption}
\usepackage{ulem}

\newcommand{\hii}{H\textsc{ii}}
\newcommand{\solmass}{M$_{\odot}$} 
\newcommand{\mewm}{$\mu$m} 
\newcommand{\kms}{kms$^{-1}$}
\newcommand{\asec}{$^{\prime\prime}$}
\newcommand{\degs}{$^{\circ}$}
 
\newcommand{\meth}{CH$_3$OH}

\newcommand{\water}{H$_2$O}
\newcommand{\thirCO}{$^{13}$CO}
\newcommand{\mma}{MM1\textit{a}}
\newcommand{\mmb}{MM1\textit{b}}

\definecolor{changes}{rgb}{0.0,0.2,1.0}

\definecolor{let}{rgb}{1.0,0.0,0.0}

\begin{document} 

   \title{Continuity of accretion from clumps to Class 0 high-mass protostars in SDC335\thanks{"The reduced interferometric data cubes and integrated intensity images in FITS format and the outflow spectra in ASCII format are only available in electronic form at the CDS via anonymous ftp to cdsarc.u-strasbg.fr (130.79.128.5) or via http://cdsweb.u-strasbg.fr/cgi-bin/qcat?J/A+A/}}
  
   \author{A. Avison \inst{1,2}\fnmsep\thanks{\email{adam.avison@manchester.ac.uk}} 
                \and G.~A. Fuller \inst{1,2,3}
                \and N. Peretto\inst{4} 
                \and A. Duarte-Cabral \inst{4}
                \and A.~L. Rosen\inst{5,6}
                \and A. Traficante\inst{7}
                \and J.~E. Pineda\inst{8}
                \and R. G\"usten \inst{9}
                \and N. Cunningham \inst{10}  }

   \institute{ Jodrell Bank Centre for Astrophysics, Department of Physics and Astronomy, School of Natural Sciences,The University of Manchester, Manchester, M13 9PL, UK
          \and
          UK ALMA Regional Centre Node
          \and
          Intituto de Astrof\'isica de Andalucia (CSIC), Glorieta de al Astronomia s/n E-18008, Granada, Spain
          \and
          School of Physics and Astronomy, Cardiff University, Queens Buildings, The Parade, Cardiff CF24 3AA, UK
          \and
          Center for Astrophysics, Harvard \& Smithsonian, 60 Garden St, Cambridge, MA 02138, USA
          \and
         NASA Einstein Fellow
          \and
          IAPS-INAF, Via Fosso del Cavaliere, 100, 00133 Rome, Italy
          \and
          Max-Planck-Institut f\"{u}r extraterrestrische Physik, Giessenbachstrasse 1, 85748 Garching, Germany
          \and
          Max-Planck-Institut f\"{u}r Radioastronomie, Auf dem H\"{u}gel 69, 53121, Bonn, Germany 
          \and 
          Institut de Radioastronomie Millimetrique (IRAM), 300 rue de la Piscine, F-38406 Saint Martin d'H\`eres, France}
   \date{}

% \abstract{}{}{}{}{} 
% 5 {} token are mandatory
 
 \abstract
{The infrared dark cloud (IRDC) SDC335.579-0.292 (hereafter, SDC335) is a massive ($\sim$5000\solmass) star-forming cloud which has been found to be globally collapsing towards one of the most massive star forming cores in the Galaxy, which is located at its centre. SDC335 is known to host three high-mass protostellar objects at early stages of their evolution and archival ALMA Cycle 0 data (at $\sim$5\asec\ resolution) indicate the presence of at least one molecular outflow in the region detected in HNC. Observations of molecular outflows from massive protostellar objects allow us to estimate the accretion rates of the protostars as well as to assess the disruptive impact that stars have on their natal clouds during their formation.}% leave it empty if necessary  
{The aim of this work is to identify and analyse the properties of the protostellar-driven molecular outflows within SDC335 and use these outflows to help refine the properties of the young massive protostars in this cloud.}
 {We imaged the molecular outflows in SDC335 using new data from the Australia Telescope Compact Array (ATCA) of SiO and Class I \meth\ maser emission (at a resolution of $\sim$3\asec) alongside observations of four CO transitions made with the Atacama Pathfinder EXperiment (APEX) and archival Atacama Large Millimeter/submillimeter Array (ALMA) CO, $^{13}$CO ($\sim$1\asec), and HNC data. We introduced a generalised argument to constrain outflow inclination angles based on observed outflow properties. We then used the properties of each outflow to infer the accretion rates on the protostellar sources driving them. These accretion properties allowed us to deduce the evolutionary characteristics of the sources. Shock-tracing SiO emission and \meth\ Class~I maser emission allowed us to locate regions of interaction between the outflows and material infalling to the central region via the filamentary arms of SDC335.} 
{We identify three molecular outflows in SDC335 -- one associated with each of the known compact \hii\ regions in the IRDC. These outflows have velocity ranges of $\sim10$\,\kms\ and temperatures of $\sim60$\,K. The two most massive sources (separated by $\sim$9000AU) have outflows with axes which are, in projection, perpendicular. A well-collimated jet-like structure with a velocity gradient of $\sim$155\kms pc$^{-1}$ is detected in the lobes of one of the outflows. 
The outflow properties show that the SDC335 protostars are in the early stages (Class 0) of their evolution, with the potential to form stars in excess of 50\solmass. The measured total accretion rate, inferred from the outflows, onto the protostars is $1.4(\pm 0.1) \times 10^{-3}$\solmass\ yr$^{-1}$, which is comparable to the total mass infall rate toward the cloud centre on parsec scales of 2.5$(\pm1.0) \times 10^{-3}$\solmass\ yr$^{-1}$, suggesting a near-continuous flow of material from cloud to core scales.

Finally, we identify multiple regions where the outflows interact with the infalling material in the cloud's six filamentary arms, creating shocked regions and pumping Class I methanol maser emission. These regions provide useful case studies for future investigations of the disruptive effect of young massive stars on their natal clouds.}
 {}


   \maketitle
%-------------------------------------------------------------------------------------------------------------------
%                             Introduction 
%-------------------------------------------------------------------------------------------------------------------
\section{Introduction}
Molecular outflows are a commonly observed feature of the star-formation process detected toward protostars that will ultimately form stars covering a wide range of main sequence masses up to spectral type B ($M_*\sim10$\solmass) and beyond. The driving force responsible for such bipolar molecular outflows is thought to be the accretion of matter onto the central protostellar object from a disk coupled with the requirement for angular momentum to be conserved in the process \citep{Konigl00,Pudritz07,Tan14,RosenOffner20}. The effects of limited resolution and obscuration impede the testing of theories of the exact physical mechanism driving these outflows observationally, be it X-wind or disk-wind driven \citep[see e.g.][]{PudritzBanerjee05, Pudritz07,Frank14}, and the extent to which magnetic fields play a role. The breadth of the scales of molecular outflows, compared to the region from which they are driven, make them an invaluable tool when testing star formation theories. For example, they can be used to indirectly probe the accretion rates of material onto forming protostars \citep{BontempsOutflow96,Ana13}.

In addition to providing information on the protostars that drive them, outflows also have an impact on their natal clouds by entraining matter as they inject energy and momentum into the surrounding interstellar medium (ISM)\citep{Offner17}. This feedback may drive turbulence and ultimately help to disrupt the cloud. These effects have implications for the number and masses of stars (i.e. the initial mass function) which can form within a single molecular cloud \citep{Ana12, Plunkett13,Krumholz14, ZhangY15,Drabek-Maunder16}. As a result, the observation of molecular outflows toward candidate high-mass protostars provides valuable information on the poorly constrained formation processes of such objects and their effects on the environments in which they form.

%--- ABOUT THE SOURCE
The Infrared Dark Cloud (IRDC) SDC335.579-0.292 \citep{Peretto09}  (hereafter, SDC335) is an increasingly well-studied star-forming region \citep{Garay02, Peretto13, Avison15} harbouring one of the most massive millimetre cores observed in the Milky Way. Seen in absorption against the mid-infrared background, SDC335 covers approximately 2.4pc at its widest extent and displays six filamentary arms, which converge at the bright infrared source at its centre. 

Using the Atacama Large Millimeter/submillimeter Array (ALMA) Cycle 0 data, \citet{Peretto13} demonstrated that the whole SDC335 cloud is in the process of global collapse and that gas (traced by N$_2$H$^+$) is flowing along the filamentary arms towards this central region. ALMA continuum data at 3mm also highlighted two mm-cores, MM1 and MM2, with $M_{core}\sim$500 and 50\solmass, respectively \citep{Peretto13}. These data also included HNC observations indicating the presence of a molecular outflow from the MM1 core, however, that outflow has not been studied prior to this work.

\defcitealias{Avison15}{Paper I}
\citet{Avison15}, (hereafter, Paper I) used radio continuum data from the Australia Telescope Compact Array (ATCA) (from 6 to 25 GHz) to reveal that the MM1 core houses two Hyper Compact \hii\ (HC\hii) regions, whilst MM2 contains a single source, which also exhibits characteristics of an HC\hii\ region. Each HC\hii\ source was coincident with a Class II methanol (\meth) maser, the pair of tracers clearly demonstrating SDC335 is in the process of forming three massive stars (each with $M_* > 9.0$\solmass, based on their calculated Lyman $\alpha$ flux). Using this constraint on the upper end of the mass range in SDC335, the authors estimate a final stellar population in SDC335 of $\sim$1400 ($M_* > 0.08$\solmass, assuming a \citealt{Kroupa02} IMF), suggesting that SDC335 could be a precursor to a massive cluster such as the Trapezium Cluster.

In this paper, we report on the molecular outflows observed in SDC335, which are likely to have been launched by the massive protostars therein. We present the observed molecular species and observations used to study the molecular outflows in SDC335 in Section 2. Section 3 describes the detected outflow properties and how they were measured. In Section 4, we discuss the measured properties and how this relates to the evolutionary status of the SDC335 high-mass protostellar objects. In Section 5, we discuss evidence for outflow-filament interactions. We present our conclusions and summarise our findings in Section 6. 
%-------------------------------------------------------------------------------------------------------------------
%                             Observations   
%-------------------------------------------------------------------------------------------------------------------
\section{Observations and ancillary data}
%--- START OBS PROPERTIES TABLE -----------
\begin{table*}
\caption[]{Instrumental setup and image product properties used within this paper.}
\begin{center}
\small
\begin{tabular}{p{2.0cm} p{1.4cm} c c c c p{1.8cm} p{1.8cm} p{1.8cm}}
\hline
\hline
 Molecule & Obs. Freq. &  Beam & MRS$^{\star}$ & Chan. Width & Sensitivity$^{\star\star}$ & \multicolumn{3}{c}{Calibrators} \\
& [GHz] & [\asec\ $\times$ \asec]& [\asec] & [MHz]/[\kms] & [mJy/bm] / [K]& Amplitude & $\phi$ & Bandpass \\
\hline
\multicolumn{9}{c}{~}\\
\multicolumn{9}{c}{Telescope: ATCA}\\
\multicolumn{9}{c}{~}\\
 SiO (1$\rightarrow$0) & 43.42385 & 3.1 $\times$ 1.3 & 23.3 & 0.125 / 0.86 & 4.3 & PKS1934-638 & PKS1646-50$^{\ddagger}$  & PKS1253-055 \\
 \meth\ (7$\rightarrow$6) & 44.06941 & 2.0 $\times$ 1.1 & 23.0 & 0.0313 / 0.21 & 9.3 & PKS1934-638 & PKS1646-50  & PKS1253-055\\
\hline 
\multicolumn{9}{c}{~}\\
\multicolumn{9}{c}{Telescope: ALMA}\\
\multicolumn{9}{c}{~}\\
 HNC (1$\rightarrow$0) & 90.66357 & 5.5 $\times$ 3.9 & 22.4 & 0.0665 / 0.22 & 14.0 & Neptune, & J1604-446 & J1517-2422 \\
& & & & & & Mercury &  & \\
 \thirCO\ (3$\rightarrow$2)$^{\dagger}$ & 330.58797 & 0.82 $\times$ 0.58 & 10.5 & 0.243 /0.22 & 23.5 & Mars, & J1650-5044$^{\ddagger}$ & J1337-1257, \\
& & & & & & Titan, &  & J1427-4206,\\
& & & & & & J1613-586 &  & J1650-5044\\
& & & & & &  & & J1924-2914\\
CO (3$\rightarrow$2)$^{\dagger}$ & 345.79599 & 0.72 $\times$ 0.54 & 11.8 & 0.254 /0.22 & 22.0 & Titan & J1650-5044, & J1427-4026,\\
& & & & &  & J1613-586 & J1517-2422\\
\hline
\multicolumn{9}{c}{~}\\
\multicolumn{9}{c}{Telescope: APEX}\\
\multicolumn{9}{c}{~}\\
CO (3$\rightarrow$2) & 345.7960 & 19.16  & - & 0.114 / 0.10 & 1.10 & - & - & - \\
CO (4$\rightarrow$3) & 461.0408 & 14.37 & - & 0.290 / 0.15 & 1.70 & - & - & - \\
CO (6$\rightarrow$5) & 691.4731 & 9.58 & - & 1.464 / 0.64 & 1.20 & - & - & - \\
CO (7$\rightarrow$6) & 806.6518 & 8.21 & - & 2.929 / 1.09 & 2.20 & - & - & - \\
\hline
\end{tabular}
\end{center}
\vspace{0.05mm}
{\tiny{\textbf{Notes:} $^{(\dagger)}$ALMA + ACA combined data.\\$^{(\star)}$ Maximum Recoverable Scale of emission, $MRS\sim0.6\frac{\lambda}{b_{min}}$, where ${b_{min}}$ is the shortest baseline in an array. \\ $^{(\ddagger)}$ PKS1646-50 and J1650-5044 are the same source under different naming conventions.\\ $^{(\star\star)}$ Unit mJy/bm for ATCA and ALMA, and K for APEX. Measured in a velocity width matching the channel width.}}
\label{Obs_props:tab}
\end{table*}
%--- END OBS PROPERITES TABLE -------------

%--- TRACERS
Molecular line emission is key to understanding the morphologies and, more importantly, the kinematics and dynamic interactions of protostellar outflows. We present the results of SiO and Class I methanol (\meth) maser observations made with ATCA toward SDC335 in combination with APEX\footnote{Atacama Pathfinder EXperiment, \citep{Gusten06}.} and archival ALMA data, observing lines of CO, \thirCO\ (ALMA only), and HNC (ALMA only), which allow us to study the outflows and their disruptive effects.
Silicon monoxide, SiO, is depleted in the gas phase of the ISM \citep{Walmsley99}, but it can be liberated from the grain mantels by shock fronts arising in molecular outflows \citep{Schilke97}, making it a useful tool for studying outflows \citep[e.g.][]{Cabral14} particularly at outflow-cloud interaction points. 

Similarly, the Class-I \meth\ maser (at 44\,GHz) is regularly used as a tracer of outflows, \citep[e.g.][]{Cyganowski09,Cyganowski11} as it is a collisionally excited maser species. Class-I \meth\ masers are frequently seen as spatially offset from the local protostellar source, unlike the radiatively pumped Class-II \meth\ maser species, lending support to the idea that the Class-I species is being excited in regions of interaction between outflowing material and the surrounding molecular material \citep{Plambeck90, Kurtz04, Vornokov10}. These lines and the observations characteristics are listed in Table \ref{Obs_props:tab}. 

Finally, CO, being ubiquitous in the ISM,  provides sufficient molecular abundance to trace the high velocity gas entrained by the outflows. Thus, it is a good tracer of the large scale molecular outflow morphology, something that is not possible when observing shock-tracing species.

\subsection{ATCA data}
The ATCA data were taken in a single 12.5 hour observing block on 3 September 2015 under the project code C3023. These observations comprised two 2 GHz continuum bands within which four 64 MHz zoom bands were placed using the CABB correlator \citep{CABBpaper}. The primary target molecular lines used from these data are SiO (1-0) and the Class-I \meth\ maser transition. 

These observations were taken with ATCA in the `750B' antenna configuration. During the data reduction, all baselines to the antenna at 6.0km from the array centre were flagged out, meaning that these data have maximum and minimum baselines of 765.3 and 61.2m, respectively. Data reduction was carried out in the software package \texttt{MIRIAD} using standard ATNF calibration and imaging strategies\footnote{See the MIRIAD user guide for more information, http://www.atnf.csiro.au/computing/software/miriad/userguide/ .}.
We note that poor atmospheric conditions at the start of observing meant that approximately four hours of data were entirely flagged from the start of the run, limiting the total observing time on source to $\sim$ 4.7 hours and setting the theoretical sensitivity to $\sim$5.64mJy/beam per 31.25kHz channel. 

\subsection{ALMA Band 7 data}
We use data taken from the ALMA archive, which forms part of the Cycle 1 observing program 2012.0.00781.S. For this work, we used the 12- and 7-m array observations at 330.64GHz and 345.874GHz, which cover the \thirCO(3-2) and CO(3-2) transitions, respectively. 

These observations comprise a 39 pointing Nyquist sampled rectangular mosaic of the central region of SDC335, which is sufficient to cover the three HC\hii\ regions observed in \citetalias{Avison15}, with the 12-m array. This mosaic has a spatial extent of 65\asec$\times$68\asec\, centred at RA = 16h30m58.550s, Dec = $-$48\degs43$^{\prime}$54.00\asec. The 7-m data had a complementary 14 point mosaic pattern covering 74.6\asec$\times$69\asec\ centred at the same position, such that the whole 12-m mosaic region was covered (and fractionally exceeded). The data were downloaded from the archive, reprocessed, and the 12- and 7-m data were combined as per ALMA recommendations\footnote{See e.g. https://casaguides.nrao.edu/index.php/M100\_Band3 for a CASA data combination tutorial.} in the data reduction package CASA \citep{CASAREF}, with a continuum subtraction applied.

For the \thirCO\ data, the array was configured with minimum and maximum baselines of 10.7 and 1284.1m, respectively \footnote{We note that in these \thirCO\ data, only ~6\% of baselines are in excess of 600m (the RMS length of all baselines being 325.65m) giving rise to a larger final synthesised beam in the combined image than would be expected from calculating the resolution using the maximum baseline.}. During the CO observations the minimum and maximum baselines were 9.1 and 437.8m, respectively. The synthesised beam sizes of these data are given in Table \ref{Obs_props:tab}.

\subsection{ALMA Band 3 data}

\begin{figure*}
\centering
\includegraphics[scale=0.6]{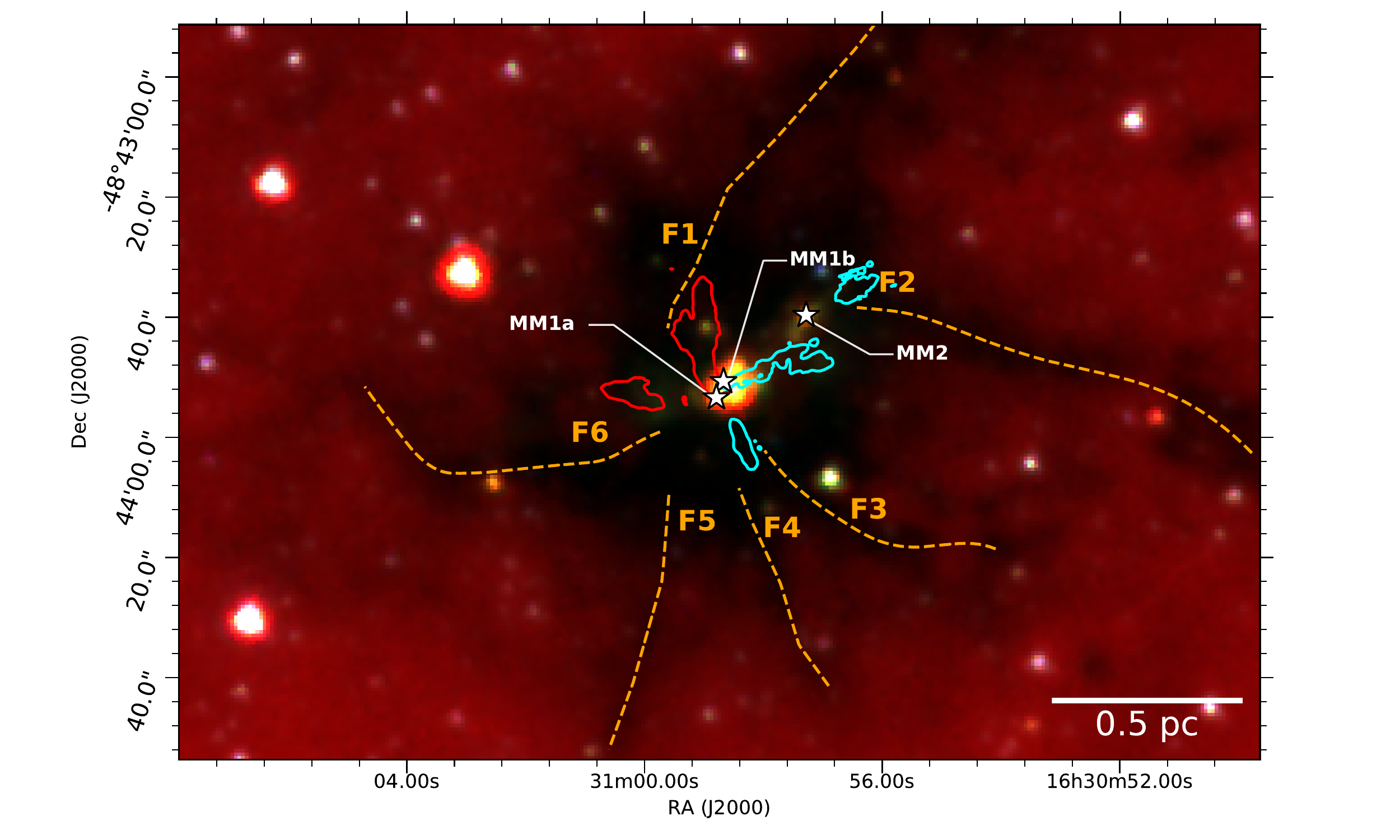}
\caption{Three colour image of SDC335 overlaid with ALMA CO contours highlighting three potential outflows within the cloud. Colour scale: \textit{Spitzer} GLIMPSE three colour image with red, green and blue using  8.0, 4.5 and 3.6\mewm\ respectively. Cyan and red contours: ALMA CO(3-2) integrated images showing the extent and morphology of the three potential outflows in the region with contours at $\sim$9\% to highlight their extent. The white $\star$ denote the locations of Class II \meth\ masers associated with the \mma, \mmb\ and MM2 compact radio cores \citepalias{Avison15}. Orange dashed lines: Nominal centroid positions of the six filaments seen in SDC335, \citep[c.f.][]{Peretto13}.}
\label{PressyFig:fig}%
\end{figure*}

ALMA Cycle 0 project 2011.0.00474.S was an 11 pointing mosaic observation at 90 GHz, covering the whole IRDC cloud seen in extinction against the mid-infrared background at 8\mewm\ from \textit{Spitzer} (see Figure \ref{PressyFig:fig} and Figure 1 of \citealt{Peretto13}). The continuum and N$_2$H$^+$ from these data were originally published in \citet{Peretto13}, though the line of interest in this current work, HNC, was not. The data were taken with 16 12-m antennas with no 7-m data taken (as this was not available with ALMA at the time). The array had maximum and minimum baselines of 18.3 and 197.7m.

\subsection{APEX CO data}
Single dish observations were made with the APEX telescope covering CO transitions (3-2), (4-3), (6-5) and (7-6). The observations of CO J=4-3 (at 461.04077 GHz) and J=3-2 (at 345.79599 GHz) were taken with the FLASH+ dual-channel receiver \citep{2014ITTST...4..588K} on 3-4 June 2013.  At the J=3-2 transition the average system temperature was $300$\,K giving a typical root mean square (RMS) of 1.1\,K in 0.1\,km/s channels.  For the J=4-3 transition, the average system temperature was 1276\,K, giving an average noise of 1.7 K in channels of a velocity width of 0.15 km/s. The data was sampled on a 3\asec\ grid and then regridded during the data reduction to a resolution of 19.2\asec\ and 14.4\asec\ full width half maximum for the J=3-2 and J=4-3 observations, respectively. The emission at the (3-2) and (4-3) transitions  shown in Figure \ref{OUTFLOW_GRID:fig}-D.

The CO\ transitions of J=6-5  (at 691.473076\,GHz)  and J=7-6 (at 806.651806\,GHz) were observed simultaneously with the seven-pixel CHAMP+\citep{2006SPIE.6275E..0NK, 2008SPIE.7020E..10G} dual channel receiver on 30 May 2013. The (7-6) transition data  were convolved to a grid with a 9.6\asec\ resolution during reduction to match the resolution of the (6-5) transition as shown in Figure \ref{OUTFLOW_GRID:fig}-E. The RMS noise level across the maps was 1.2\,K in 0.64 km/s channels and 2.2\,K in 1.1 km/s channels for the lower and higher frequency transitions, respectively. 

%-------------------------------------------------------------------------------------------------------------------
%                             Outflows   
%-------------------------------------------------------------------------------------------------------------------

%------------------------------ START Pressy image ------------------------------------------
%------------------------------- END Pressy image --------------------------------------------

\section{Outflow identification}
\label{Results:sec}
To identify potential outflows within our data, a continuum subtracted data cube was created  for each molecular line. These cubes have a velocity range from $-$100 to 0 \kms\, which is sufficient to cover the whole kinematic range of the observed species given the $V_{lsr}$ of the target. The MM1 core has a $V_{lsr}$ of $-$46.6\kms\ and for MM2 $V_{lsr}$ = $-$46.5\kms\ \citep{Peretto13}. The $V_{lsr}$ of the compact radio cores observed by \citetalias{Avison15} are consistent with those of the mm-cores based on their associated Class II (6.7-GHz) methanol masers, \mma\ $V_{\rm{CH_3OH}}$=$-$48.0 to $-$45.0\kms, \mmb\ $V_{\rm{CH_3OH}}$=$-$56.0 to $-$50.0\kms\ and MM2 $V_{\rm{CH_3OH}}$=$-$51.0 to $-$43.0\kms \citep{MMB330to345},  and the observed modulus offsets between molecular gas and peak maser emission of  \noindent $\sim$ 3 to 4 \kms\ \citep{Szymczak07, Pandian09, GreenMcClure11}. As such, we adopted a $V_{lsr}$ of $-$46.6\kms\ for \mma\ and \mmb\ and $-$46.5\kms\ for MM2 in line with the mm-core values.

Using these cubes, the velocity ranges of each identified outflow for each molecular species were found by visual inspection and are listed in Table \ref{OutflowCalc:tab}, (defined as the absolute offset from $V_{lsr}$ to the last channel with a 3$\sigma$ detection). Emission close to the $V_{lsr}$ of the system is complex, particularly in the ALMA CO maps. As such, we implemented velocity range limits for our interpretation of the structures in SDC335. For all species, except CO, we excluded emission $\pm5$\kms from the $V_{lsr}$. For calculations of outflow properties with CO we use a larger velocity offset from the $V_{lsr}$ as described in Section \ref{DerivProps:sec}. The velocity ranges used for each outflow by species are given in Table \ref{OutflowCalc:tab}. We then generated integrated intensity maps over these velocity ranges from which we measured the observed spatial extent and opening angles of each outflow wing. 

\begin{figure*}
\centering
\includegraphics[scale=0.64]{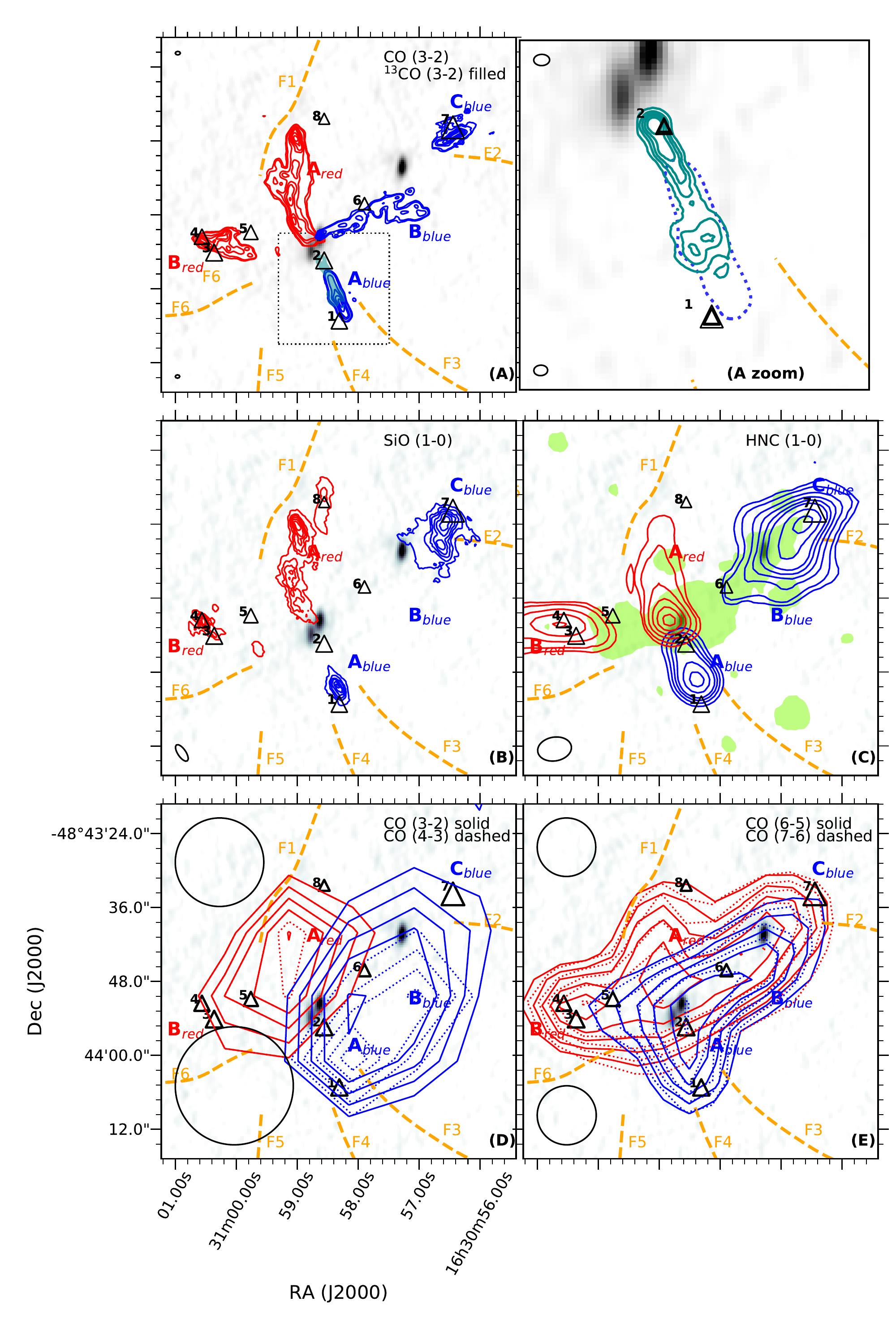}
\caption{\small{Outflows in SDC335. Each panel includes the following data of the SDC335 region. \textit{Greyscale}: 8 GHz continuum emission from \citetalias{Avison15}. \textit{Orange dashed lines}: Filament centroids as per Figure \ref{PressyFig:fig}. \textit{Triangles}: Class-I \meth\ maser positions (c.f. Figure \ref{MaserCO:fig} and Table \ref{Maser:tab}). The ellipse in the bottom left of each plot and at top left in panels \textit{(A)}, \textit{(D)} \& \textit{(E)} gives observed beam shape, which are the synthesised beam widths for interferometric and the HPBW for single dish observations, respectively. Beam shape values are given in Table \ref{Obs_props:tab}.  Each panel shows the integrated red- and blue-shifted emission from the detected outflows as red and blue contours, respectively over the velocity ranges listed in Table \ref{OutflowCalc:tab}. Panel \textit{(A)} shows the outflows in ALMA CO(3-2), as unfilled contours, and \thirCO(3-2), as filled contours, the dashed box around outflow A$_{blue}$ denotes the region shown in the adjacent panel. Panel \textit{(A zoom)}, Zoom in of region bordered by the dashed box in Panel \textit{(A)}, here the \thirCO(3-2) emission is shown as the solid contour at the 10, 30, 45, 60, 75 and 90\% of the peak  integrated emission. The 10\% contour of CO(3-2) is shown as the dotted contour for comparison to Panel \textit{(A)}. Panel \textit{(B)} ATCA SiO(1-0) emission. Panel  \textit{(C)} ALMA HNC(1-0) emission, this panel also includes as the filled green contour the 4.5\mewm\ emission above above 12.5 MJy/sr for the EGO observed in SDC335 \citep{Cyganowski09}. \textit{(D)} APEX detected CO(4-3), as solid contours, and CO(3-2), as dashed contours, emission at  at 50, 60, 70, 80 and 99\% of the APEX data peak emission at each frequency. \textit{(E)} APEX detected CO(6-5), as solid contours, and CO(7-6), as dashed contours, emission at  at 50, 60, 70, 80 and 99\% of the APEX data peak emission at each frequency. For panels \textit{(A)} \& \textit{(B)} the contour levels are at  10, 30, 45, 60, 75 and 90\% of the peak emission per outflow, for panel \textit{(C)} the contour levels are at (5), 10, 20, 30, 40, 50, 70 and 90\% of the peak emission per outflow (5\% contour for red lobe only, to emphasise the curvature noted in the text).}}
\label{OUTFLOW_GRID:fig}%
\end{figure*}

\begin{figure}
\centering
\includegraphics[scale=0.55]{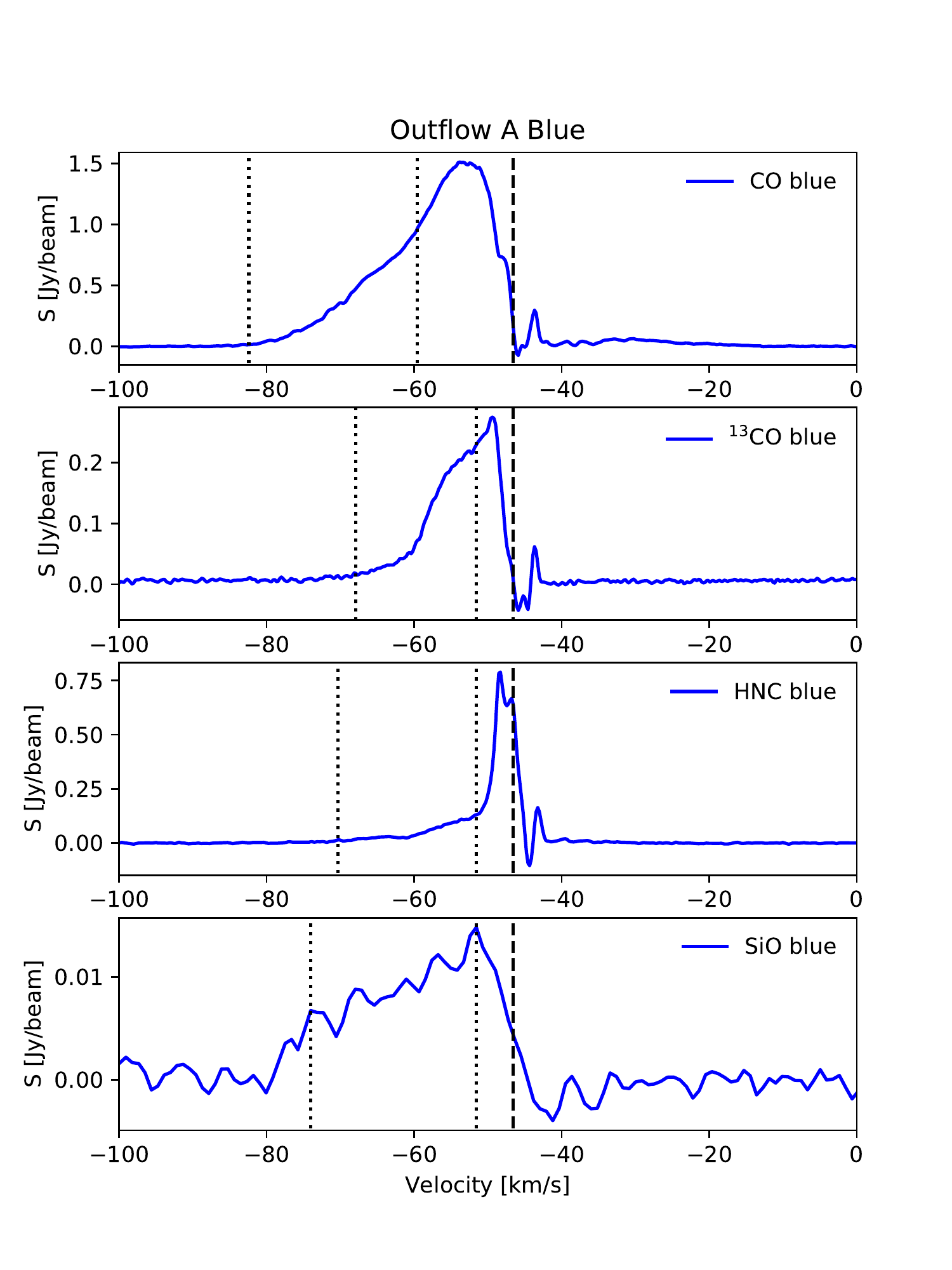}
\caption{Averaged spectra measured toward the blue lobe of outflow A for species CO, \thirCO, HNC and SiO. The spectra are measured within the outer contour of the integrated intensity maps from Figure \ref{OUTFLOW_GRID:fig} for each species. The vertical dashed line gives the $V_{lsr}$ of the target and the vertical dotted lines mark the range in velocity from Table \ref{OutflowCalc:tab}, based on which the integrated intensity maps were produced as were the CO outflow
property calculations. The chosen limits are described in the main text. }
\label{OUTA_SPEC_blue:fig}%
\end{figure}

\begin{figure}
\centering
\includegraphics[scale=0.55]{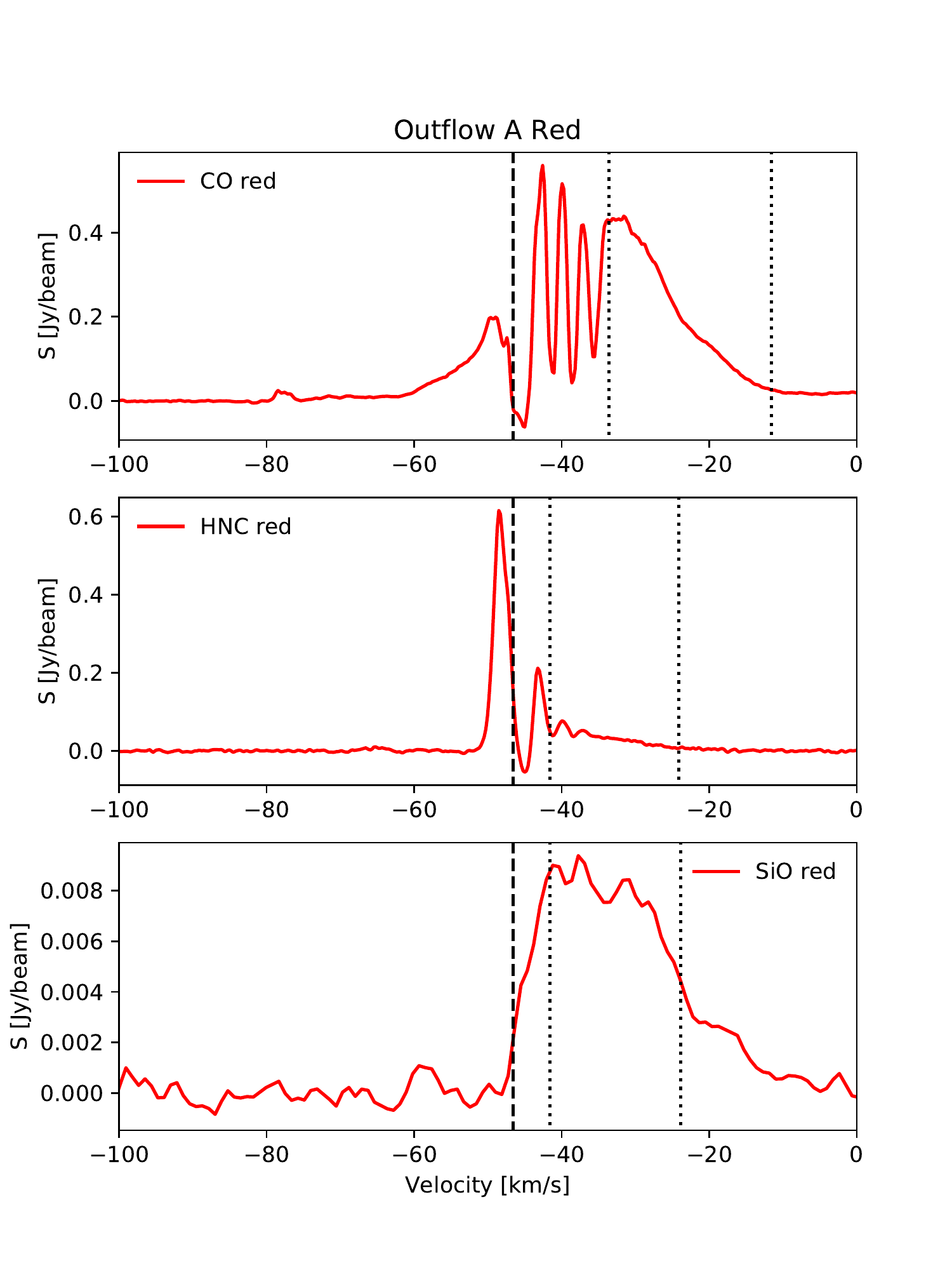}
\caption{Averaged spectra measured toward the red lobe of outflow A for species CO,  HNC, and SiO. Measured and line markings as per Figure \ref{OUTA_SPEC_blue:fig}.  }
\label{OUTA_SPEC_red:fig}%
\end{figure}

\begin{figure}
\centering
\includegraphics[scale=0.55]{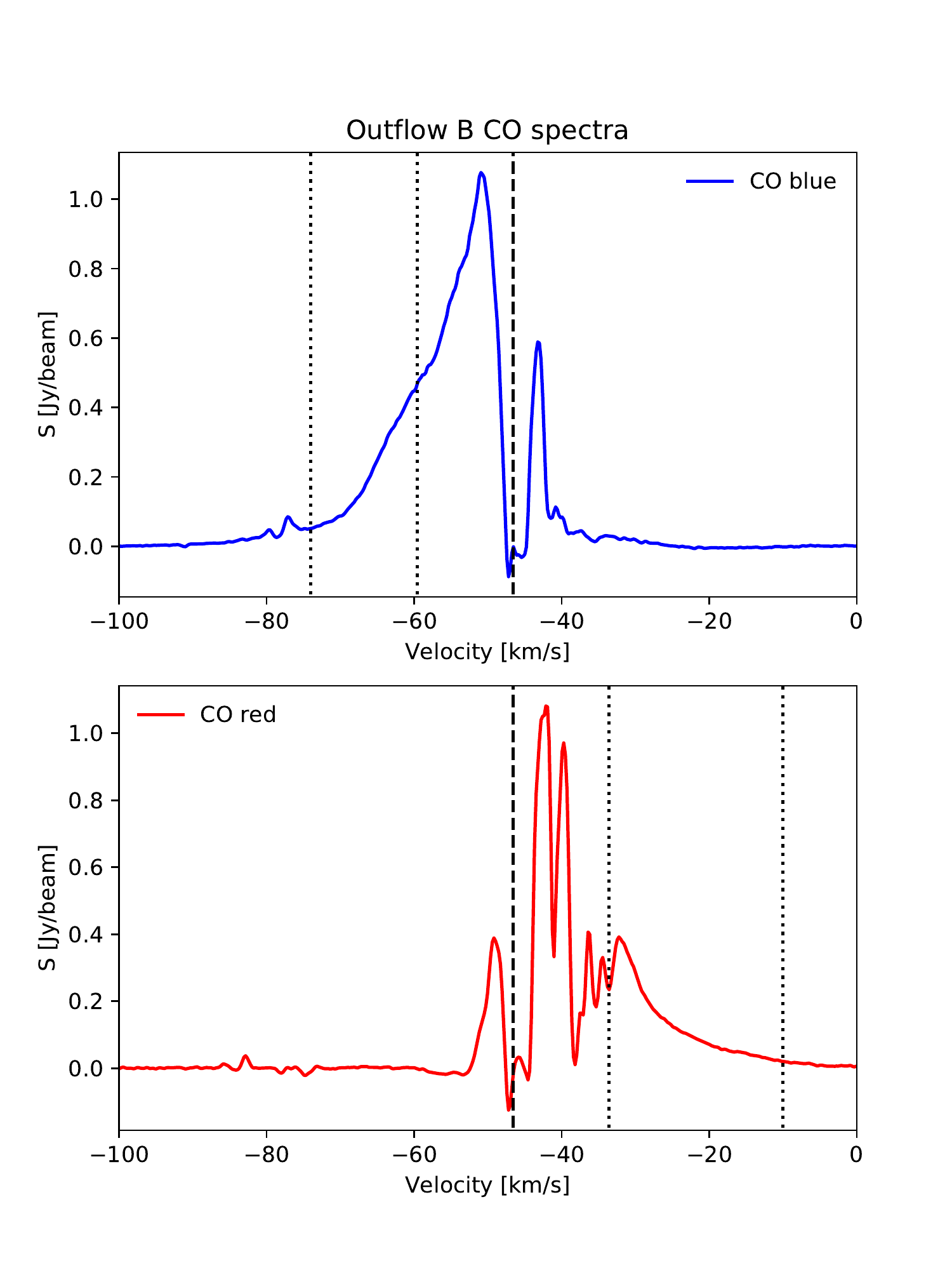}
\caption{Averaged spectra measured toward the blue and red lobes of outflow B for species CO. Measured and line markings as per Figure \ref{OUTA_SPEC_blue:fig}.  }
\label{OUTB_SPEC:fig}%
\end{figure}

\begin{figure}
\centering
\includegraphics[scale=0.55]{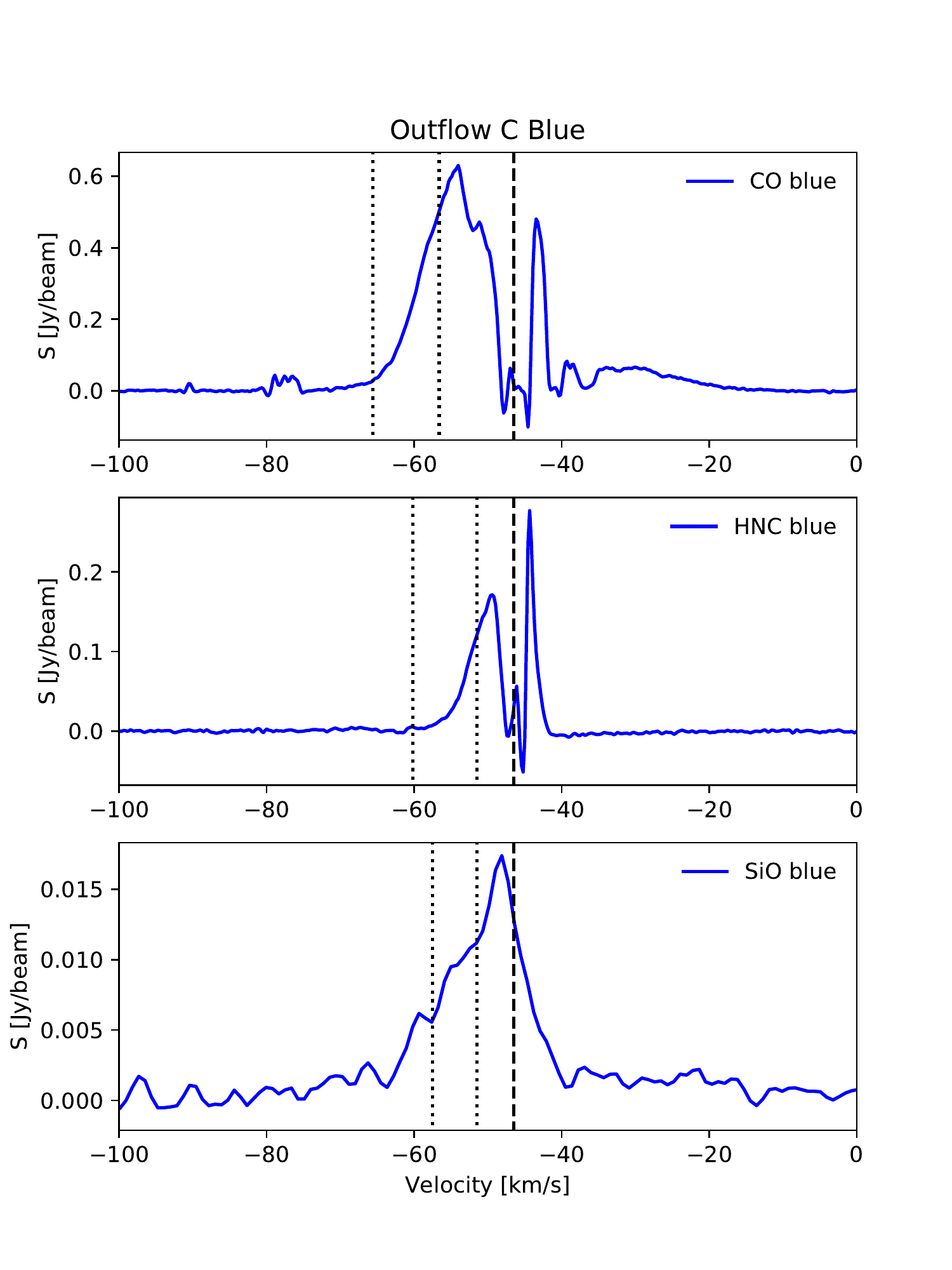}
\caption{Averaged spectra measured toward the blue (and only detected) lobe of outflow C for species CO, HNC, and SiO. Measured and line markings as per Figure \ref{OUTA_SPEC_blue:fig}.  }
\label{OUTC_SPEC:fig}%
\end{figure}

Within the multiple datasets presented in this paper, there are three distinct molecular outflows observed within SDC335. We label these A, B, and C and we describe their respective morphologies, along with the identification of their likely progenitor protostars in Section \ref{OutDescript:sec}. Figure \ref{PressyFig:fig} shows components of each outflow superimposed on a \textit{Spitzer} IRAC three colour image of the SDC335 IRDC to highlight their spatial extent within the larger scale cloud. Figure \ref{OUTFLOW_GRID:fig} presents the morphology of each outflow as observed in CO, \thirCO\, HNC (from ALMA), SiO (from ATCA), and CO from APEX. In Figures \ref{OUTA_SPEC_blue:fig} to \ref{OUTC_SPEC:fig}, we present the spectra of each detected line from each outflow within the interferometric data. We also present spectra over the same velocity range as the outflows (-100 to 0 \kms) at the peak position of each of the three HC\hii\ at the CO, \thirCO, HNC, and SiO tunings in Appendix \ref{HCHII_SPEC:app} as Figures \ref{MM1a_SPEC:fig} to  \ref{MM2_SPEC:fig}.

%------------------------------ START Outflow Grid Plot ------------------------------------------
%------------------------------- END Outflow Grid Plot --------------------------------------------

%---------------------------- START OUTFLOW SPECTRA -----------------------------------
%---------- A BLUE

%---------- A RED

%---------- B

%---------- C

%---------------------------------------
%  OUTFLOW Description
%---------------------------------------
\subsection{Outflow descriptions}
\label{OutDescript:sec}
\subsubsection{A: North-south molecular outflow}
The most apparent of the three outflows, namely, outflow A, runs approximately from north to south through the MM1\footnote{A note on nomenclature: within this paper, we use MM1 to indicate the millimetre emission core from \citet{Peretto13}, while \mma\ and \mmb\ to refer to the smaller radio emission/HC\hii\ regions from \citetalias{Avison15}, which are both encompassed within the MM1 mm-core.} core. The outflow is detected in all tracers, SiO, HNC, CO (both ALMA and APEX), and \thirCO\, (see Figure \ref{OUTFLOW_GRID:fig}). However, the red northern component of this outflow is not detected in the \thirCO\ data. This is consistent with the trend seen throughout all the presented data, which shows that red-shifted emission is found to be weaker within each outflow. We attribute this to the geometry of the system, which is caused by the obscuring effect of self-absorption by the intervening material. %As observed in ALMA CO the red wing covers a velocity range from 

It is not immediately clear to determine which of the two HC\hii\  regions within MM1 is the progenitor of outflow A. As noted in \citetalias{Avison15}, the area of maximal overlap between the two wings of the HNC outflow (Figure \ref{OUTFLOW_GRID:fig}, panel C) is coincident with the position of the \mma\ compact radio source. At a higher resolution, the morphology of the red-shifted emission seems to show a slight curvature toward \mmb\ within the central core region, as seen in CO (3-2) and SiO (Figures \ref{OUTFLOW_GRID:fig}-A and \ref{OUTFLOW_GRID:fig}-B), but this could be due to interactions between the two outflows A and B (\S \ref{DescB:sec}) originating from this region. Since the blue-shifted emission is elongated linearly with the major axis of emission aligned toward the \mma\ core, we treat \mma\ as the progenitor of this outflow. We measure an average length-to-width ratio for this outflow, at the 10\% of peak integrated intensity contour in Figure \ref{OUTFLOW_GRID:fig}-A, using both the red and blue lobes of 3.5.

\subsubsection{Jet-like properties in outflow A}
\begin{figure}
\centering
\includegraphics[scale=0.5]{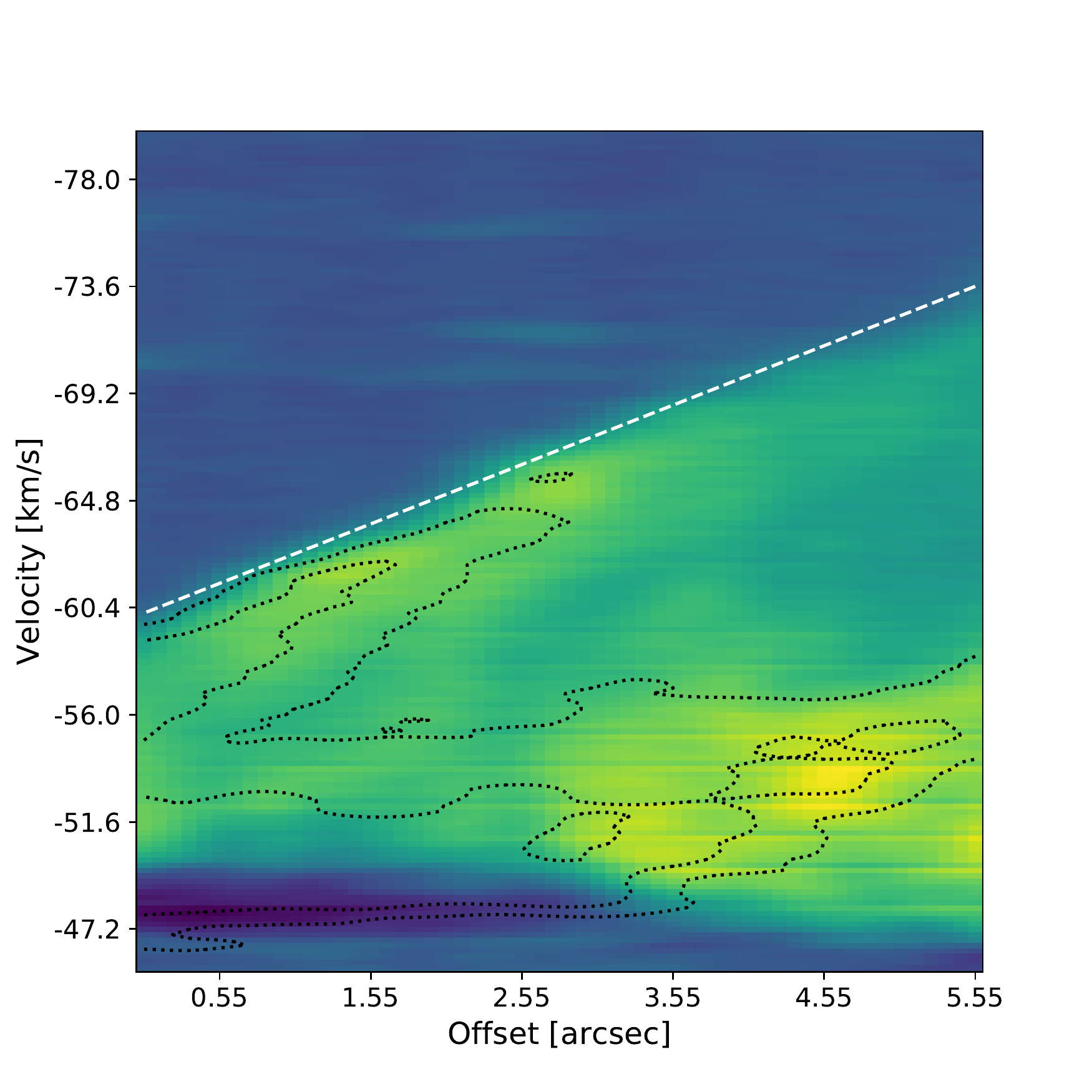}
\caption{Position velocity digram along the  A$_{Blue}$ outflow axis. The PV data covers a physical size of 0.09pc offset from the driving source (\mma) and stopping before the `knot-like' feature described in the text. \textit{Colourscale:} ALMA CO(3-2) data. \textit{Contours:} ALMA $^{13}$CO(3-2) data. The white dashed line shows the the observed velocity gradient of $\sim$155 \kms\ pc$^{-1}$.}
\label{PVJet:fig}%
\end{figure}

An inspection of the ALMA CO and \thirCO\ data shows that the blue lobe of outflow A is highly collimated and exhibits a high velocity gradient. The length-to-width ratio of the ALMA CO and \thirCO\ contour at the 10\% of peak integrated intensity contours in Figure \ref{OUTFLOW_GRID:fig}-A$_{zoom}$ are 2.8 and 3.4, respectively, matching the fiducial lower bounds of collimation of a jet structure ($\sim$ 3 e.g. \citealt{Arceetal06}) and comparable to other observed molecular jet sources, such as HH211 \citep{Gueth99} and Outflow C in IRAS 05358$+$3543 \citep{Beuther02}. Figure \ref{PVJet:fig} gives a position velocity (PV) diagram along the outflow axis as observed in CO outward from the driving source and covering a physical scale of $\sim 0.09$pc. We see that there is a clear linear velocity gradient within (highlighted by the white dashed line) from both CO emission (colour scale) and \thirCO\ (displayed as a 5 and 10$\sigma$ contours). From the CO data, we calculate a velocity gradient of 155 \kms\ pc$^{-1}$. This value is comparable to molecular outflow velocities seen in other high mass sources, such as K3-50A \citep{Klaassen13}.
%------------------------------ START  Jet PV ------------------------------------------
%------------------------------- END Jet PV --------------------------------------------

\subsubsection{B: East-west molecular outflow}
\label{DescB:sec}
Outflow B runs approximately east to west through the MM1 core and is seen primarily in the CO (both ALMA and APEX), with the red lobe seen in both HNC and SiO. The APEX CO data (at all four frequencies) is dominated by an EW component (Figure \ref{OUTFLOW_GRID:fig}, D and E). The known extended green object (EGO) in SDC335 \citep{Cyganowski08} is orientated in this same east-west (EW) direction (green filled contour in Figure \ref{OUTFLOW_GRID:fig}-C). The 4.5\mewm\ emission responsible for EGO features is thought to be created predominantly by shocked H$_2$ in outflowing material \citep[][and references therein]{Cyganowski11}. Outflow B displays an average length-to-width ratio as measured in the ALMA CO data of 3.0.

From the ALMA CO(3-2) data (Fig. \ref{OUTFLOW_GRID:fig}-A), the morphology of this outflow would suggest that \mmb\ is the likely progenitor. Interestingly it is approximately perpendicular, as projected on the sky, to the A outflow. This suggests, if \mma\ and \mmb\ represent individual protostars, that their respective axes of rotation are likely to have very different orientations. Reasons for this misalignment in SDC335 are not clear from these data, but a possible scenario is considered in Section \ref{RosenSims:sec}.

\subsubsection{C: NW (SE) molecular outflow}
Outflow C is located to the north-west (NW) of the MM2 core. Detected in interferometric data in CO, HNC, and SiO, the molecular emission appears to be coming from a single outflow lobe, with its counterpart undetected, likely due to it having a similar velocity to the systemic velocity of SDC335 and, therefore, remaining in the excluded velocity range. This is supported by the presence of red-shifted emission in the APEX CO data that suggests the presence of a counterpart. This outflow hosts the brightest shock-tracing SiO emission of the three and the strongest Class-I \meth\ maser detected here. Due to the lack of a counterpart red lobe, we do not attempt to measure an average length-to-width ratio for this outflow.

\subsubsection{Considering other possible outflows}
The detection of the three outflows described above does not preclude the existence of additional outflows within SDC335. Using the ALMA band 7 CO data (the dataset with both the highest sensitivity and spatial resolution), we see that close to the $V_{lsr}$ of the system,  the potential for additional structures still remains. However, the current data in this region suffers from two issues arising from interferometric artefacts: firstly, the bright CO emission and, therefore, the associated interferometric side-lobe artefacts are complex at these velocities; and secondly, owing to a lack of zero-spacing or single dish data, the ALMA image suffers from negative `bowling' that comes from resolved-out extended emission in the source\footnote{see e.g. \citet{Braun85} and Chapter 8 \citet{SynthesisImaging} for an explanation of bowling as different to interferometric side-lobes.}, thereby further complicating the assessment of features at these velocities. These issues, combined with a lack of evidence for other outflows in our data at other frequencies, lead us to disregard some tentative structures which may appear to be additional outflows.

%--- START OUTFLOW PROPERITES Vrange TABLE -----------
\begin{table}
\caption[]{Velocity ranges used in calculating outflow properties and integrated intensity maps by observed species from ALMA and ATCA observed transitions. }
\begin{center}
\begin{tabular}{c c c c}
\hline
\hline
Outflow & Molecular & Lobe & $V_{range}$\\
 & species & & [\kms] \\
\hline
\multirow{7}{*}{A}  & \multirow{2}{*}{CO} & Blue & $-$59.6 - $-$82.4 \\
 & & Red & $-$11.6 - $-$33.6\\ 
&\thirCO & Blue & $-$51.6 - $-$67.9   \\ 
 &\multirow{2}{*}{SiO} & Blue & $-$51.6 - $-$74.0  \\ 
 & & Red & $-$23.9 - $-$41.6    \\ 
  &\multirow{2}{*}{HNC} & Blue & $-$51.6 - $-$70.3  \\ 
 & & Red & $-$24.1 - $-$41.6\\
\hline
\multirow{2}{*}{B}  & \multirow{2}{*}{CO} & Blue &  $-$59.6 - $-$74.0 \\
 & & Red & $-$10.0 - $-$33.6\\ 
 %&SiO & Red & YY  \\ 
% &HNC & Red & yy  \\ 
\hline
\multirow{3}{*}{C}  & CO & Blue & $-$56.6 - $-$65.6 \\
 &SiO & Blue & $-$51.5 - $-$60.2 \\ 
 &HNC & Blue &$-$51.5 - $-$57.5  \\ 
 \hline
\end{tabular}
\end{center}
\label{OutflowCalc:tab}
\end{table}
%--- END OUTFLOW Vrange TABLE -----------

%---------------------------------------
%  OUTFLOW PROPERTIES 
%---------------------------------------
\subsection{Outflow properties}
%----- START MAIN Outflow properties TABLE -----------------------------
\begin{table*}
\caption[]{Measured and derived outflow properties. }
\begin{center}
\begin{tabular}{c c c l c | c c | c c c | c}
\hline
\hline
 & & & & & \multicolumn{2}{ c |}{Measured} & \multicolumn{3}{c |}{Corrected for inclination$^{\dagger\dagger}$} & \\
Outflow & V$_{lsr}$$^{\dagger}$ &Inclination$^{\dagger\dagger}$ & Molecular & Wing & $l$ & $\mid$V$_{max}$$\mid$ & $l$ & $\mid$V$_{max}$$\mid$ & t$_{dyn}$ & Associated\\
 & [\kms] & Angle [\degs] & Species &  & [pc] & [\kms] &  [pc] & [\kms] & [10$^3$ years] & Masers$^{\ast}$ \\
\hline
\multirow{7}{*}{A} &\multirow{7}{*}{$-$46.6} & \multirow{7}{*}{53-76} & CO (3-2)& Red & 0.30 & 35.0 & 0.34 & 87.6 & 3.8 & \multirow{7}{*}{1, 2, 8} \\
 & &  &$^{13}$CO (3-2) & Red & - & - & - & -  &\\ 
 & &  &SiO (1-0) & Red & 0.29 &  22.7 & 0.32 &  56.7 & 5.5 &\\
 & &  &HNC (1-0) & Red & 0.30 &  22.5 & 0.33  & 56.3 & 5.8 &\\
 & &  &CO( 3-2)& Blue & 0.16 & 35.8 & 0.18 & 89.5 & 1.9 &\\
 & &  &$ ^{13}$CO (3-2) & Blue & 0.12 & 21.3 & 0.14 & 53.2 & 2.6 & \\
 & &  &SiO (1-0)& Blue & 0.17 & 27.4 & 0.19 & 68.5 & 2.7&\\
 & &  &HNC (1-0)& Blue & 0.16 & 23.7 & 0.18 & 59.5 & 3.0 &\\
\hline
\multirow{2}{*}{B} &\multirow{2}{*}{$-46.6$} &\multirow{2}{*}{59-79} & CO& Red & 0.30 & 36.6 & 0.33 & 110.9 & 2.9 & \multirow{2}{*}{3, 4, 5, 6} \\ 
 &  & &CO & Blue & 0.25 & 27.4 & 0.25 & 83.1& 3.2 &  \\
\hline
\multirow{3}{*}{C} &\multirow{3}{*}{$-$46.5} &\multirow{3}{*}{57.3} & CO (3-2)& Blue & 0.19 & 19.1 & 0.22 & 35.4 & 6.1 & \multirow{3}{*}{7} \\ 
 &  & &SiO (1-0) & Blue & 0.15  & 13.7 & 0.18 & 25.3 & 7.1 & \\
 &  & &HNC (1-0) & Blue & 0.16 & 11.0 & 0.19 & 20.4 & 9.2 &\\
\hline
\end{tabular}
\end{center}
\vspace{0.05mm}
{\tiny{\textbf{Notes:}\\$^{(\dagger)}$ $V_{lsr}$ is local standard of rest velocity of the outflow driving source, see text for details.\\ $^{(\dagger\dagger)}$ For outflows A and B, we present values of $l$, $|V_{max}$| and t$_{dyn}$ calculated using the average value of the correction factor function of the inclination angle range in column 3, the values of the correction factors are given in Table \ref{CorrectionFactors:tab}. For outflow C, we use the single value of 57.3\degs. See text for definitions and inclination correction factors.\\ $^{(\ast)}$ Associated Class I methanol masers as numbered in Table \ref{Maser:tab} and Figure \ref{MaserCO:fig}}.\\}
\label{outflowprop:tab}
\end{table*}
%----- END MAIN Outflow properties TABLE --------------------------------

\subsubsection{Inclination angles}
\label{incangle:sec}
\begin{figure*}[]
    \centering
    \begin{subfigure}[t]{0.5\textwidth}
        \centering
        \includegraphics[scale=0.6, trim = 220 160 160 120, clip]{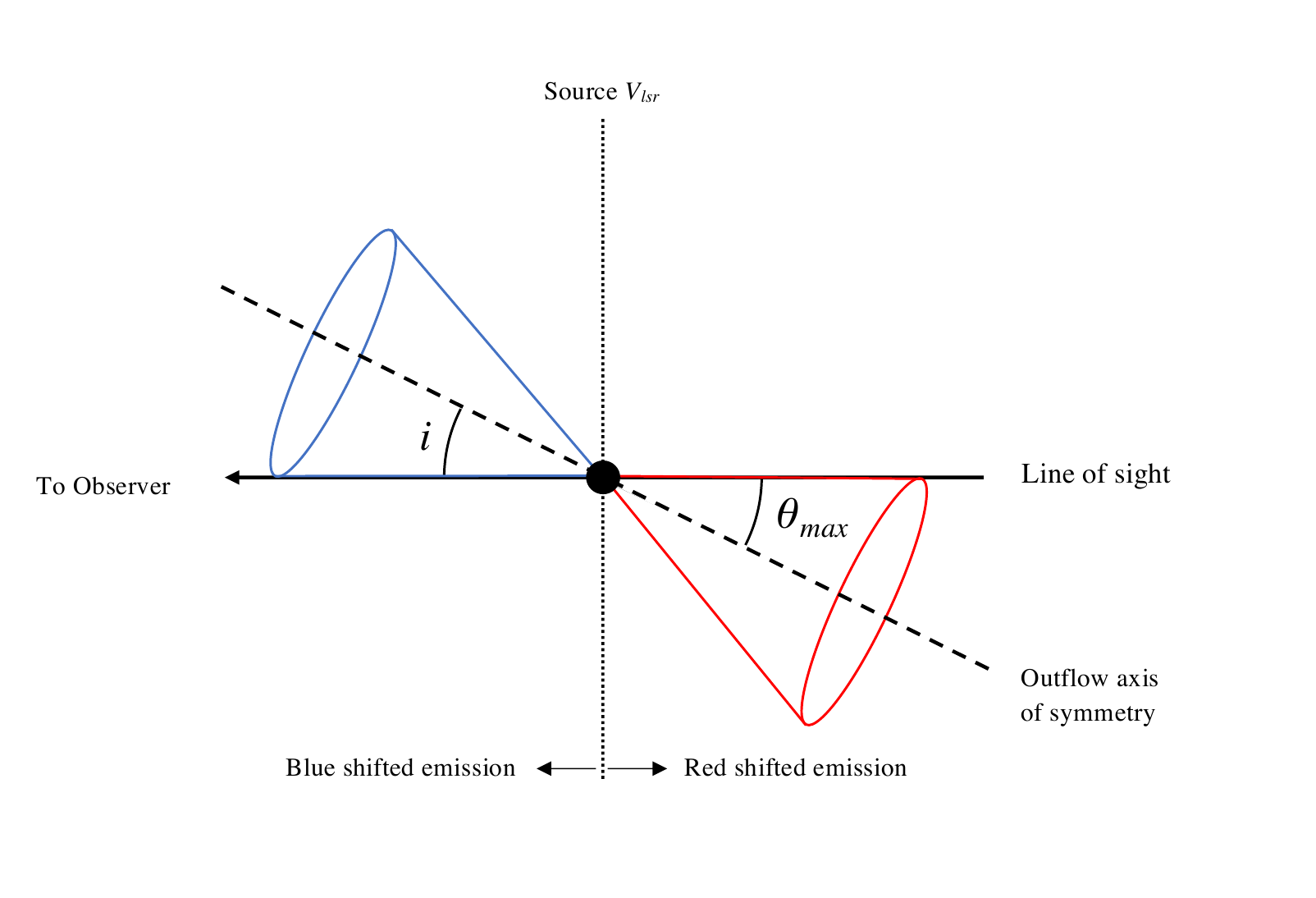}
        \caption{Limiting case A: $i \geq \theta_{max}$, here $i\sim\theta_{max}$.}
        \label{iGTtheta:fig}
    \end{subfigure}%
    ~ 
    \begin{subfigure}[t]{0.5\textwidth}
        \centering
        \includegraphics[scale=0.6, trim = 220 160 160 120, clip]{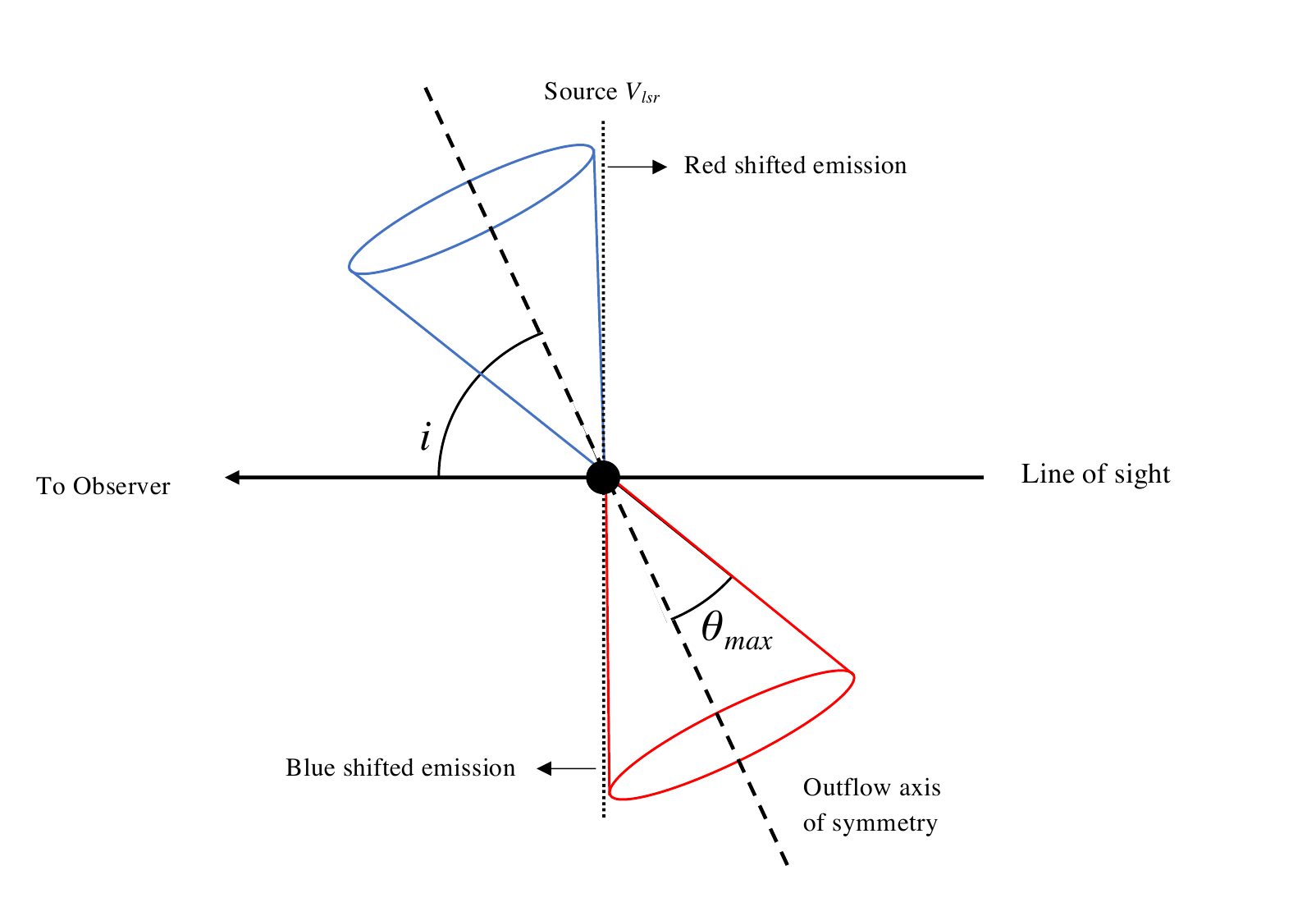}
        \caption{Limiting case B: $i \leq 90-\theta_{max}$, here $i\sim90-\theta_{max}$.}
        \label{iGT90mintheta:fig}
    \end{subfigure}
\caption{Two limiting cases of the \citet{CabritBertout86} `case 2' outflow morphology ascribed to outflows A and B. The \citet{CabritBertout86} `case 3' outflow  occurs when $i$ is sufficiently large to make the rear of the blue lobe and front of the red lobe cross the dotted $V_{lsr}$ mode in, for example, shown here in (b). }
\label{outflowIncExplain:fig}%
\end{figure*}

To estimate the properties of the detected outflows, we must consider their inclination to the plane of the sky, measured as the angle $i,$ with $i=0$\degs\ indicating that the axis of the outflow is oriented along the observer's line of sight and $i=90$\degs\ indicating that it is perpendicular to the light of sight. To assess this, we used the ALMA CO integrated intensity maps (shown in Figure \ref{OUTFLOW_GRID:fig}-A) as this dataset has the highest spatial resolution and signal-to-noise ratio.  In the first instance, we assume each outflow has a symmetric biconical geometry, following \citet{CabritBertout86}, and we define the opening angle, $\theta_{max}$, as the angle between the cone's axis of symmetry and the outer emission contour at 3$\sigma$ (see Figure \ref{outflowIncExplain:fig}). 

%------------------------------ START TOY OUT FIG ------------------------------------------
%------------------------------- END TOY OUT FIG --------------------------------------------

Next, we compare the measured $\theta_{max}$, the observed positions and morphology of the red and blue outflow lobes on the plane of the sky to the model outflows described by \citet{CabritBertout86}. The work of \citet{CabritBertout86} includes four model `cases' of differing inclination angle, $i$, and the opening angle (see their Table 1) to describe the morphology of the red and blue lobes an observer would expect to see on the sky for each given case. In brief\footnote{For more details, please refer to the original text in \citet{CabritBertout86}.} the four cases are as follows:
%\LEt{ AA style avoids the use of bullets in the text. Please keep the separation into paragraphs and simply remove the bullet dash.}

Case 1 describes an outflow with both a red-shifted and a blue-shifted lobe on the line of sight, namely, with blue in front of a red lobe. The inclination angle conditions for this case are $i < \theta_{max}$ and $i \leq 90^{\circ} - \theta_{max}$. Observationally, this yields an outflowing red-blue lobe pair, which overlap spatially on the line of sight.

Case 2 describes a case with two observed outflow lobes, again one red-shifted and one blue-shifted, but with the inclination angle conditions of this case as $i \geq \theta_{max}$ and $i \leq 90^{\circ} -\theta_{max}$ (labelled A and B in Figure \ref{outflowIncExplain:fig}). Observationally there will be no point in the plane of the sky where the red and blue lobes spatially (or kinematically) overlap.

Case 3 has outflow lobes which are do not overlap spatially (as per case 2) but with the emission from each lobe in the pair spanning the source, $V_{lsr}$, resulting in mixed red-and-blue-shifted emission from each lobe. The \citet{CabritBertout86} model here requires inclination angles of values of $i \geq \theta_{max}$ and $i > 90^{\circ} -\theta_{max}$.

Case 4 requires very large opening angle ($\theta_{max}$) and inclination angles which obey these conditions, $i < \theta_{max}$ and $i > 90^{\circ} -\theta_{max}$. This results in lobes that overlap spatially (as per Case 1) and also span the $V_{lsr}$ and, as such, exhibit both red-shifted and blue-shifted emission (as per Case 3).

%\noindent 

We note that for the purposes of the current study, only cases 2 and 3 are relevant here.
Outflow A has two spatially and kinematically non-overlapping lobes in the plane of the sky. This is indicative of a \citet{CabritBertout86} case 2 source.  We find that the opening angle, $\theta_{max}$, of the blue-shifted part of outflow A is $\sim 14^{\circ}$, putting loose limits on $i$ of between 14 and 76\degs. Similarly, outflow B has two non-spatially or kinematically overlapping lobes, again indicating a \citet{CabritBertout86} case 2 source. We measure $\theta_{max}$ as between $\sim$9.4 and 12.6\degs, and use the average value of 11\degs\ in calculations giving $i$ between 11 and 79\degs. 

Determining the inclination of outflow C is complicated by the fact that only one lobe is reliably detected to the north-west of MM2. We could assume that C is a \citet{CabritBertout86} case 3 source ( $i\geq \theta_{max}$ and i $> 90^{\circ}$) given that the observed velocity range of the one observed lobe spans the $V_{lsr}$ of MM2. This would give an inclination angle of $>77^{\circ}$ given our measured $\theta_{max}$ of 13.4$^{\circ}$. However, as the red lobe is not reliably detected in this source, we cannot confirm that the red lobe would similarly demonstrate an emission spanning the $V_{lsr}$ and, therefore, we use, instead, the commonly adopted statistical average value of $i =57.3^{\circ}$. This average inclination angle, $\langle \theta \rangle$ is calculated as:
\begin{equation} 
\langle \theta \rangle \int_{0}^{\pi} \rm{d}\Omega =  \int_{\theta = 0}^{\pi/2} \int_{0}^{\pi}\theta \sin \theta \rm{d}\Omega\rm{d}\theta = 1\rm{rad} \simeq 57.3^{\circ} \label{AvAngle:eqn}.\end{equation}. 
%\LEt{ I\ recommend placing the equation here in the main text instead of as a footnote.}

\subsubsection{Further consideration of the outflow inclination angles for outflows A and B.}
\label{angConsider:sec}
Based solely on the \citet{CabritBertout86} models the potential inclination angles for outflows A and B cover ranges of 62$^{\circ}$ and 68$^{\circ}$, respectively. Such broad ranges introduce significantly different correction factors when deriving outflow properties. These correction factors range between 1.0 and 5.2 to correct the measured length, velocity, and momentum, with between 1.0 and 27.5 for energy and 0.2 and 27.0 for momentum flux (see Section \ref{DerivProps:sec} for a description of these derived properties and their respective correction factors). To further limit the possible range of inclination angles, we now outline an additional consideration, beyond the \citet{CabritBertout86} models, based on the observed length-to-width ratios ($LWR_O$) and the observed distance from the driving source to the maximum width ($x_O$) of the observed outflows. We present in Appendix \ref{incAngle:app} the full geometric derivation of this morphology and our associated results.

Outflows A and B have observed length-to-width ratios of 3.5 and 3.0, respectively (average of both red and blue lobes for each). Given also these outflows' observed opening angles, and assuming the biconical outflow morphology of \citet{CabritBertout86}, we find that it is not possible to recover such observed length-to-width ratios with no external influence on the outflow. With their respective opening angles and a bicone morphology, the maximum recoverable $LWR_O$ for each outflow reaches an asymptote at angles of 76 and 79\degs\, and with values of $LWR_O$ of 2.1 and 2.6 for A and B, respectively. This is due to the fact that under the biconical outflow morphology the observed length, $L_O$, is also the point along the outflow of maximum observed width. 

In observations and modelling of molecular outflows, the idealised biconical morphology is seldom, if ever, seen (see e.g. \citealt[and references therein]{Frank14}). This is due to the influence of the environment on the outflow. In particular, modelling the effects of the collapsing clumps, precession, turbulence, and magnetic fields \citep[see e.g.][]{Rosen20} show that the outflow morphologies are narrower along the outflow axis than the bicone morphology would suggest. Given these effects and the fact that we cannot reproduce the observed \textit{LWR} for outflows A and B using the biconical morphology, we instead propose an alternate morphology to limit the inclination angle ranges of these outflows.

%\sout{Next we take a second simple morphology where}\sout{Again assuming no external influence on the morphology of the outflows,}
Our proposed `pencil-like' morphology allows the widest point can be at some arbitrary point, $x$, along the length (as is observed for outflows A and B).  Using this second morphology (as detailed in Appendix \ref{incAngle:app} and Figure \ref{NewOutflow:fig}) we find inclination angle ranges of between 53 and 76\degs\ for outflow A and 59 and 79\degs\ for outflow B which can provide the observed $LWR_O$. 

An alternate scenario by which the observed $LWR_O$ for outflows A and B could be achieved would consist of observing these outflows at low inclination angles where the observed emission primarily arises from the outflow cavity rather than the outflow edge (i.e. Figure \ref{iGTtheta:fig}, as opposed to Figure \ref{iGT90mintheta:fig} or similar for our alternate morphology). In such a case, the observed length-to-width ratios would require shaping of the outflow cavity away from symmetry about the outflow direction by the surrounding medium, compressing and elongating the cross-section of the cavity in a single dimension. In the case of SDC335, to observe two outflows from driving sources separated by $<$9000 AU which display similar levels of compression of their outflow cavities but at orthogonal directions seems implausible.

We therefore consider it more likely that the outflows are inclined closer to the plane of the sky than along the line of sight. As such, we consider the lower bounds from the \citet{CabritBertout86} models as extreme values and base our analysis instead on values ranging from those allowed by the geometric arguments presented in Appendix \ref{incAngle:app} of 53 to 76\degs\ for outflow A and 59 to 79\degs\ for outflow B. 

Table \ref{CorrectionFactors:tab} provides the correction factors used for each outflow to correct the derived values to create the inclination angle corrected values given in Tables \ref{outflowprop:tab} and \ref{CO_outflowprop:tab}. For outflow C we use the correction factor at 57.3\degs\ and for outflows A and B we used average value of the correction function over the angle range given in column 3 of Table \ref{outflowprop:tab}. For completeness, we also include in Table \ref{CorrectionFactors:tab} the correction factors which would apply if the inclination angle ranges were based solely on the \citet{CabritBertout86} cases (given in italics) and in, Appendix B, a brief discussion of the implications of using these correction factors.

The proper assignment of inclination angles to observed outflows is a difficult task and has significant implications on derived properties. With current leading observatories, for example, ALMA, providing a wealth of observed molecular outflows, it is timely for investigation to be set into a robust method of assigning these angles. We look forward to the possibility that the use of an alternate morphology in this work will raise further discussions of this matter.

%----------- CORRECTION FACTORS TABLE ---------------
\begin{table*}
\caption[]{Table of factors used corrected observed outflow properties to account for the inclination angle, $i$. The average value of the function over the angle range is given for outflows A and B and for outflow C the value at 57.3\degs\ is presented. Additionally, for outflows A and B, we also provide in italics the correction factors applicable discounting the considerations made in Section \ref{angConsider:sec} and Appendix \ref{incAngle:app}. }
\begin{center}
\begin{tabular}{c c c c c}
\hline
\hline
Derived & Correction &  Average in range 53-76\degs\ \textit{(14-76\degs)}& Average in range 59-79\degs\ \textit{(11-79\degs)}& Value at 57.3\degs\ \\
property & factor & (Outflow A) & (Outflow B) & (Outflow C)\\
\hline
Length, $l$ & $\frac{1}{\sin i}$ & 1.12  \textit{(1.71)} &  1.08 \textit{(1.81)} & 1.19\\
Velocity, $|V_{max}|$, & \multirow{2}{*}{$\frac{1}{\cos i}$} & \multirow{2}{*}{2.50 \textit{(1.71)}}   &  \multirow{2}{*}{3.03  \textit{(1.81)}}  & \multirow{2}{*}{1.85}\\
Momentum, $P$ & & & &\\
Energy, $E$ & $\frac{1}{\cos ^2 i}$& 6.69 \textit{(3.47)} & 9.97  \textit{(4.17)}& 3.43\\
\\Momentum Flux, $F$ & $\frac{\sin i}{\cos ^2 i}$& 6.16  \textit{(2.86)} & 9.45 \textit{(3.56)} & 2.88\\
\hline
\end{tabular}
\end{center}
\label{CorrectionFactors:tab}
\end{table*}
%-------------------------------------------------------------------------

\subsubsection{Temperature estimation}
\label{Radex:sec}
\begin{figure}
\centering
\includegraphics[scale=0.5]{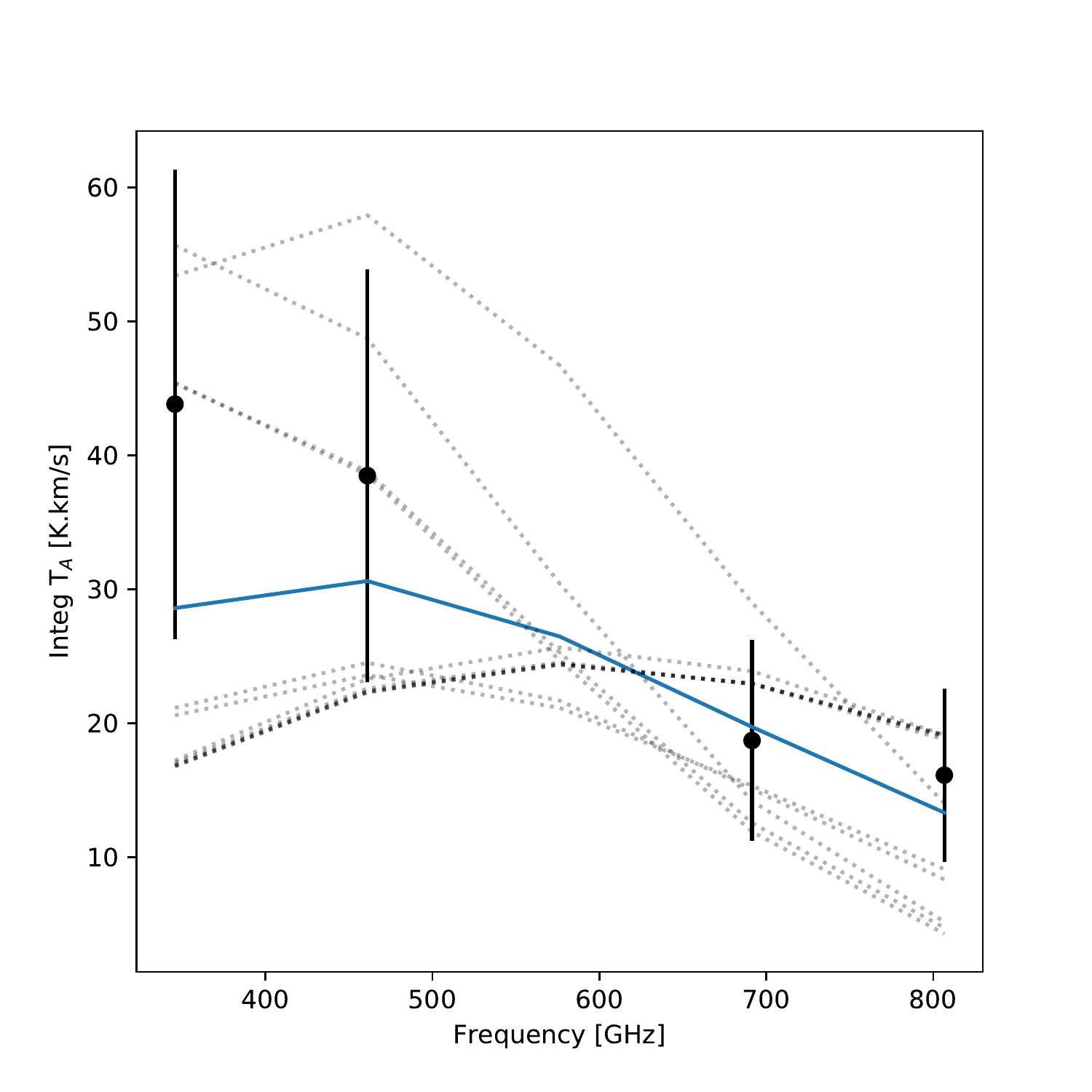}
\caption{Example spectral line energy distribution used for fitting the temperature toward the blue lobe of outflow A. The data (black circles with errorbars) are at CO transitions (3-2), (4-3), (6-5), and (7-6). The model fits (dotted and solid line) from RADEX also include CO (5-4). The dotted line represents the best fit ($\chi^2=1\pm0.05$) SLEDs from the RADEX grid. The solid line represents the average value for these best fit models in (for this example) SLED. This fit gives T$_{kin}$ = 66K.}
\label{SLED:fig}
\end{figure}

The four observed transitions of CO within the APEX data allow us to estimate the kinematic temperature of the outflowing gas in SDC335. To achieve this, we first regridded the data from each observation to the poorest spatial and velocity resolution, those being APEX CO (3-2) spatially and APEX CO (7-6) in velocity. We then created integrated intensity maps at 5\,\kms\ intervals from each cube. For each observed outflow we measured the integrated intensity (in K\,\kms) at two positions along the outflow, creating a spectral line energy distribution (SLED) at each position. We then followed the approach of \citet{Liu18} by generating a large grid over temperature, number and column density with the radiative transfer code RADEX \citep{VandertakRADEX} and searching for the best-fit modelled intensities compared to our observed SLED intensities to give a temperature at a given position. The best fit being a $\chi^2\sim1$ following the $\chi^2$ approach of \citet{vanderTak2000}. We then take all models with $\chi^2=1\pm0.05$ and use the average properties of these models. Figure \ref{SLED:fig} gives an example of the APEX CO intensities and best fit model at a given position and velocity along outflow A. Using this technique we find $T_{kin}$ values of 62/53K, 54/61K and 55K for the blue and red components of outflows A, B and C respectively, (see Table \ref{CO_outflowprop:tab}).

%------------------------------ START SLED FIG ------------------------------------------
%------------------------------- END SLED FIG --------------------------------------------

\subsubsection{Derived properties}
\label{DerivProps:sec}

Using the inclination angles discussed in Section \ref{angConsider:sec},  we were able to calculate inclination corrected physical properties for each outflow. For each observed species, we provide in Table \ref{outflowprop:tab} both the measured and inclination angle corrected outflow lobe length, $l$ and $|V_{max}|$. We use the inclination-corrected values to calculate a dynamical time, $t_{dyn}$, for each outflow lobe. We define these properties and their inclination angle correction factors following \citet[and references therein]{Cunningham16}, as follows:
%\LEt{ AA avoids the use of bullets in the main text. Please consider reworking these back into the paragraph. Make sure not to start a sentence with a symbol }

The lobe length, $l$, is the maximum distance from the progenitor peak to the $3\sigma$ emission contour over the wings velocity range. The corrected length is calculated as $l_{corr}=l_{measured}/\sin(i)$ for inclination angle $i$. The absolute velocity difference, $|V_{max}|$, is defined as the difference between the outflow progenitor's $V_{lsr}$ and the final velocity channel in the image cube to show $3\sigma$ emission from the outflow, with a correction for $i$ is calculated as $|V_{max_{corr}}|=|V_{max_{meas}}|/\cos(i)$. Finally, $t_{dyn}$, the dynamical age of the outflow is calculated as $t_{dyn}=l_{corr}/|V_{max_{corr}}|$.

For the ALMA CO observations, we also provide in Table \ref{CO_outflowprop:tab} the mass, momentum, and energy of each outflow lobe, following the methods of \citet{Ana12}.  As part of this analysis, we note that, as may be expected for a high mass star forming molecular cloud, emission close to the systemic velocity of the cloud becomes increasingly complex, with features of self absorption and missing large scale flux in the interferometric data, which can lead to uncertainties in values derived from emission observed within these velocity ranges. In order to avoid basing our findings on velocity ranges within the data effected by self-absorption or imaging artefacts, we exclude data close to the systemic velocity in the following way. 

For the MM1 core, we take the spectrum of CO (3-2) data from APEX over the whole SDC335 velocity range. We fit a Gaussian to this spectrum, centred at the $V_{lsr}$ and exclude emission from channels $\pm2.5\times$ the variance, $\sigma$, of this Gaussian fit. The fitted $\sigma$ was found to be 5.2\kms. The factor of 2.5 was chosen to ensure that all self-absorption and missing flux features in the interferometric data were excluded for the MM1 core outflows (A and B) as can be seen in Figures \ref{OUTA_SPEC_blue:fig}, \ref{OUTA_SPEC_red:fig}, and \ref{OUTB_SPEC:fig}. Excluding these velocity channels means our derived properties are lower limits for outflows A and B.

We repeat the same process for the MM2 core, taking the APEX CO (3-2) spectrum and fitting a Gaussian. Here, $\sigma$ is found to be 6.7\kms. The velocity range of the observed outflow lobe C from source MM2 is very close to the $V_{lsr}$ of MM2 that to exclude emission from $\pm2.5\sigma$ velocity channels would remove too much emission to provide a realistic determination of the values. Instead we use  $\pm1.5\sigma$ sufficient to avoid self absorption artefacts in the CO data and retain sufficient data to get a meaningful result. The velocity ranges used for calculating outflow properties and generating the integrated intensity images of each outflow lobe shown in Figure \ref{OUTFLOW_GRID:fig} are given in Table \ref{OutflowCalc:tab}.
 
Although we have other molecular species in the current data we do not carry out the following calculations with the data for these reasons.  In particular, SiO is a shock-tracing species and therefore cannot reliably trace the bulk outflow properties since it will be enhanced in regions of shocks and less abundant elsewhere; HNC is currently a poorly studied outflow tracer therefore it is not clear whether or not (like CO) it traces the bulk outflow properties and further studies of this species in multiple targets would be required to assess its nature in outflows, which is beyond the scope of this paper. 

For our calculations with the CO ALMA data, we define mass within the outflow, $M_{out}$ in units of \solmass, as:

\begin{equation}
\begin{split}
M_{out}=\left(\frac{10^3}{M_{\odot}}\right)\left(\frac{8Q\pi k \nu^2}{A_{ul}hg_uc^3} \right)\exp\left(\frac{E_{ul}}{T}+\frac{h\nu}{kT}\right)\\\times\left(\frac{\mu/N_{A}}{M_{\odot}}\right)D^2  \sum \limits_{x,y,v}  \tau_{corr} T_{mb} \rm{d}V
\end{split}
\label{Mout:eqn}
,\end{equation}

\noindent where $\mu$ is the mean molecular mass, $N_{A}$ the Avagadro's constant, $Q$ the partition function, and $A_{ul}$ the Einstein coefficient. $E_{ul}$ and $g_{u}$  are the energy of the transition in K and the statistical weight of the upper energy level (2J+1), respectively, and D is this distance to SDC335 (3.25kpc). The value given by $ \sum \limits_{x,y,v}{T_{mb} \rm{d}}V$ is equivalent to the integrated intensity and is calculated from the data cubes within a polygon region around each lobe. The sum is made per pixel, $x, y$, per velocity channel, $v$, over the outflows respective velocity range for all pixels with an intensity greater than or equal to three times the RMS per channel. Within this sum the data units are converted from the native units Jy/beam to Kelvin. The factor $\tau_{corr}$ gives the optical depth correction factor of the form $\tau_{corr}=\frac{\tau}{1-e^{-\tau}}$ and d$V$ gives the channel velocity width in \kms.

We calculate the mass both with a lower bound temperature $T$ of 20K and a literature value for $\tau_{corr}$ of 3.5 \citep{CabritBertout92} as well as the temperature values recovered from our best fit radiative transfer models from \S \ref{Radex:sec} (Table \ref{CO_outflowprop:tab}) with a $\tau_{corr}$ factor of 8.2 based upon our estimated $^{12}$CO optical depth. To estimate an optical depth for $^{12}$CO in the outflows, we use the \thirCO\ to $^{12}$CO line ratio observed in the blue wing of outflow A. Following the procedure of \citet{Myers83}, we compare the \thirCO/$^{12}$CO line ratios observed at multiple points (at peak emission) along the outflow to the ratio of optical depths given by $\frac{1-e^{-\tau_{^{13}CO}}}{1-e^{-\tau_{^{12}CO}}} $ and use the assumption that $\tau_{^{12}CO}$ is equal to $f\times \tau_{^{13}CO}$. Here, $f$ is the isotopic abundance ratio, in the range $f=50-100$ typical of star-forming regions \citep[e.g.][]{Pineda10,Szucs14}. Given this range of $f$ we can derive numerically a range of optical depths for \thirCO\ and thus attain a corresponding range of  $\tau_{^{12}CO}$ values. We find a range of $\tau_{^{12}CO}$ = 2.4 and 23.5  and use the median value of 8.2. These are comparable to typical values seen in CO outflows e.g. \citep{CabritBertout92} of  1 to 8.9 with a median of 3.5.  

Momentum and energy then follow as:

\begin{equation}
P_{CO}=\sum \limits_{x,y,v} M_{x,y,v} \Delta V
\end{equation}

\noindent and

\begin{equation}
E_{CO}=\frac{1}{2}\sum \limits_{x,y,v} M_{x,y,v} \Delta V^2
,\end{equation}

\noindent where $\Delta V$ is the absolute velocity difference between the source $V_{lsr}$ and the given channel in the sum. $M_{x,y,v}$ is the outflow mass value for only a single voxel ($x,y,v$), which follows from Equation \ref{Mout:eqn}. Finally, we calculate the momentum flux $F_{CO}$ within annular regions projected out from the central source which is defined (following \citealt[][]{BontempsOutflow96}, their equation 1 and \citealt{Ana13}), as:

\begin{equation}
F_{CO}=\sum \limits_{x,y,v} \left( \frac{\Omega \tau_{corr} T_{mb} \rm{d}V \Delta V^2}{{\rm{d}}r}\right)
\label{FCO:eqn}
,\end{equation}

\noindent where d$r$ is the width of the annular ring in the sum and $\Omega$ is the same constant as the value preceding the integral in Equation \ref{Mout:eqn}. %\LEt{ Please use either Equation spelled out in full or Eq. consistently throughout.} 

To take into account the assumed inclination angles, the values of $P_{CO}$, $E_{CO}$, and $F_{CO}$ are corrected by the factors $1/\cos(i)$, $1/\cos ^2(i)$ and $\sin(i)/\cos ^2(i)$, respectively, with their specific values for each outflow and outflow property given in Table \ref{CorrectionFactors:tab}. The error in inclination angle is the most significant contribution to the errors on these values. The derived outflow property values are given in Table \ref{CO_outflowprop:tab}.

Our derived values are comparable to those from both observed and theoretical studies of low- to high-mass protostellar objects \citep[e.g.][]{Zhang05,Yildiz12,Cunningham16,vanKempen16,Cyganowski17,Staff18}, with the SDC335 outflows A and B typically having amongst the highest values compared to published sources.  The values for outflow C are somewhat lower and likely affected by the limited velocity range we were able to use for this feature. These results confirm SDC335 protostellar sources are indeed high-mass protostars. When compared to \citet{Maud15}, the SDC335 sources appear to have a similar range of momentum fluxes and energies, but significantly lower mass and momentum. This difference is likely due to comparing single dish \citep{Maud15}  and interferometric (this work) data with the potential for multiple spatially unresolved outflows in the single dish data boosting values within the \citet{Maud15} sample.

\section{Infall, accretion, and the evolutionary status of SDC335}
\subsection{Cloud infall and protostellar accretion rates}
\label{infall:sec}
Outflow momentum flux (Equation \ref{FCO:eqn}) is a proxy for the protostellar mass accretion rate, $\dot M_{acc}$, since the driving force of outflows is thought to result from the conservation of angular momentum within an accreting system \citep{Konigl00,Pudritz07,Tan14}. A theoretical relation between $F_{CO}$ and $\dot M_{acc}$ equates the two values scaled by a material entrainment efficiency, $f_{ent},$ and the protostellar mass ejection rate and wind speed, $\dot M_{w}, v_w$  \citep[see their Equation 4]{Ana13}. Following this form of the $F_{CO}$ to $\dot M_{acc}$ relation and assuming the same material entrainment efficiency and wind properties as these authors have done ($f_{ent} = 0.5$, $\dot M_{w}/\dot M_{acc} = 0.15$ and $v_w = 40$\kms), we derive the mass accretion rates given in Table \ref{CO_outflowprop:tab}.

For outflows A and B, the $\dot M_{acc}$ derived from our $F_{CO}$ values are in the range of 8.7 - 64.8 $\times 10^{-5}$\solmass\ yr$^{-1}$ , at a fixed temperature of $T=20$K,  and 11.5 - 85.4 $\times 10^{-5}$\solmass\ yr$^{-1}$, for our derived temperatures ($T=53 - 62$K). For outflow C, values are lower at $\dot M_{acc}$ is 0.2 $\times 10^{-5}$\solmass\ yr$^{-1}$ for $T=20$K and 0.3 $\times 10^{-5}$\solmass\ yr$^{-1}$ at the derived temperature of $T=55$K. 

The values for outflows A and B are at the higher end of inferred accretion rates from low-mass protostars, $\dot M_{acc} \sim 10^{-5}$\solmass\ yr$^{-1}$, \citep[e.g.][]{McKeeTan03,HosokawaOmukai09, HosokawaYorkeOmukai10} and are more typical of those inferred for high-mass protostars, $10^{-4} - 10^{-3}$\solmass\ yr$^{-1}$ \citep[see e.g.][]{Zhang05,Fuller05,Beuther13,Ana13,Rosen16,Goddi18, Rosen19}, from both observational and theoretical works reported within the literature. The value for  $\dot M_{acc}$ of outflow C is very low and likely a consequence of the limited velocity range over which we are able to measure this outflow's properties. Indeed, as noted in Section \ref{OutDescript:sec}, the driving sources for all three of the observed outflows are known to be high-mass protostars owing to their association with 6.7GHz methanol masers, so the derived $\dot M_{acc}$ for outflow C can act as a lower limit. 

We now consider the total derived mass accretion, $\dot M_{tot. acc}$, within SDC335. This value is calculated as the sum of all the derived $\dot M_{acc}$ values presented in Table \ref{CO_outflowprop:tab}, giving a $\dot M_{tot. acc}$ of 1.4 ($\pm$ 0.1) $\times 10^{-3}$\solmass\ yr$^{-1}$. With this value, an interesting comparison can be made with the derived mass infall rate, $\dot M_{inf}$, for the whole SDC335 cloud. \citet{Peretto13} calculated a value of  $\dot M_{inf} \simeq 2.5 (\pm 1.0) \times 10^{-3}$\solmass\ yr$^{-1}$. These two values are comparable within the quoted errors, which would imply a near 100\% efficiency of infalling material on cloud scales ($\sim$few pc) being funnelled onto the accretion disk scale ($<<0.1$pc) and driving the outflows. We note, however, that this does not imply a 100\% accretion of the inflowing material into the protostar. A near continuous flow of material from large (cloud/filament) to smaller (clump/core) scales would be significant as it suggests that each core is acting as a sink within the cloud drawing material from all scales and, as such, would have implications for high-mass star forming models. The protostellar accretion rate is a characteristic of the whole forming cluster under competitive accretion models \citep{Bonnell01, Bonnell04, Bonnell07} but under core accretion models \citep{McKeeTan02, McKeeTan03, Tan14}, this is set by the properties of bound cores. It would appear that this result favours the former set of models.

Another interesting implication of these inferred accretion rates is that the three known protostellar sources in SDC335 would appear to be drawing the majority of the material onto themselves, suggesting that there is a limited amount of material available to start forming any additional protostars within the system. In ALMA observations of the IRDC G28.34 P1, \citet{Zhang15} found a lack of low-mass protostars compared to the expected number from a typical IMF. These authors concluded, after discounting migration of low-mass stars into the centre of the cluster, that the most likely scenario for this under-abundance was because the low-mass stars are yet to form. If this is true then SDC335 may hold an explanation, in that early onset high-mass star formation dominates the inflowing material budget depriving lower mass proto or pre-stellar cores from gaining mass and starting formation.

As the inclination angles of the observed outflow are the dominant factor in the error budget of these calculations, an additional study of the protostars in SDC335 is required to better constrain the outflow properties. For example, testing for the presence of rotation axes, indicative of accretion disks, within the sources would help further this investigation.

\subsection{Evolutionary Status of SDC335}
\label{EvoStat:sec}
The works of \citet{Ana13}, \citet{Maud15} and \citet{vanKempen16}, for instance, have shown there is a linear (in log-space) relation between the molecular outflow momentum flux (and therefore $\dot M_{acc}$) and source luminosity in protostellar sources from low to high-masses. Recent numerical works by \citet{Rosen20} show this same trend.

%------------------------------ START FOUT LBOL FIG ------------------------------------------
\begin{figure*}
\centering
\includegraphics[scale=0.6]{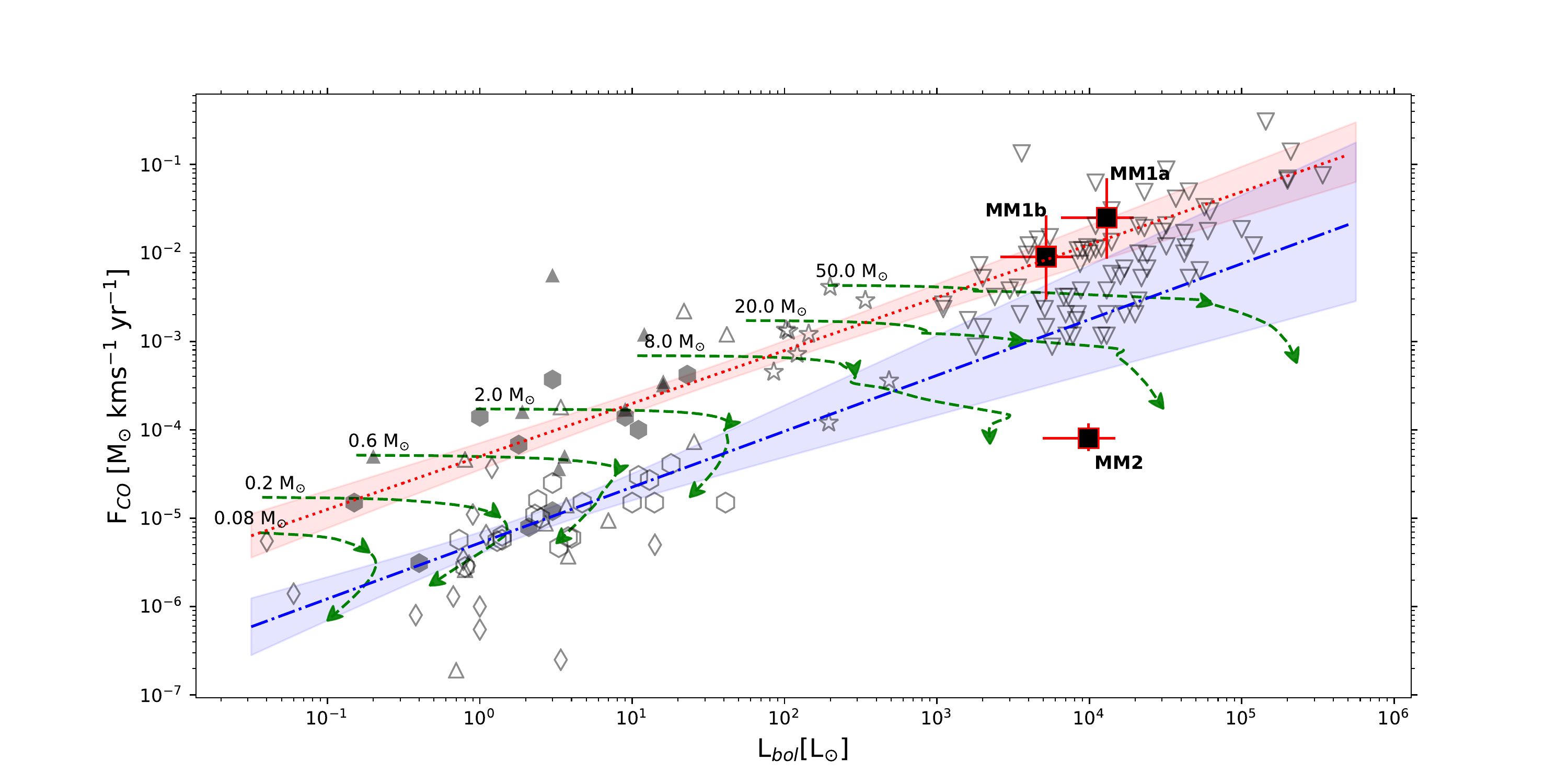}
\caption{Outflow momentum flux, $F_{CO}$,  as a function of powering source bolometric luminosity $L_{bol}$. Data for outflows A, B, and C in SDC335 are indicated by the filled squares and associated errorbars. The origin of the errorbars are discussed in the main text. The filled and empty hexagons are the Class 0 and Class I low-mass protostars from \citet{BontempsOutflow96}. The diamonds are low-mass Class I sources from observation of Ophiuchus  \citet{vanderMarel13}. Filled and empty triangles are literature values for Class 0 and I objects used by \citet[see their Table E.1]{vanderMarel13}.  The stars denote the \citet{Ana13} high-mass Class 0 analogues and the triangles are outflows from high-mass protostellar objects observed from the RMS survey, \citet{Maud15}. The green dashed tracks with arrow heads are the evolutionary tracks for decreasing or intermittent accretion from \citet{Ana13}. The arrow heads represent the point at which a protostar of a given mass (given at the end of the arrow) has accreted 50\% and 90\% of their envelope mass. Red dotted line is the best fit to the plotted Class 0 sources extended to high luminosity with the red shaded region given the 1-$\sigma$ error margin for this fit. The blue dashed-dot line and shaded region are the same but for the best fit to the plotted Class I sources.}
\label{Fout_Lbol:fig}
\end{figure*}
%------------------------------- END FOUT LBOL FIG --------------------------------------------

Figure \ref{Fout_Lbol:fig} plots the calculated outflow momentum flux as a function of source bolometric luminosity, $L_{bol}$, for the protostars in SDC335 and other sources, both low and high-mass, from the literature. The sources from the literature are low-mass Class 0 and 1 sources using values from \citet[and references therein]{BontempsOutflow96} (both classes), observed and literature Class I sources from \citealt[][their Table 4 and E.1, respectively]{vanderMarel13}.  High-mass sources in the plot are Class 0 analogues from \citet{Ana13} and high-mass protostellar sources from \citet{Maud15}.

The $L_{bol}$ values for SDC335 are those derived in \citetalias{Avison15}, and we divide the total luminosity of MM1 (1.82$\times 10^4$ $L_{\odot}$) between \mma\ and \mmb\ in a 2.5:1 ratio (based on their respective optically thin radio emission flux ratios, see \citetalias{Avison15} Table 5) yielding values of 1.3$\times 10^4$ and 0.51$\times 10^4$~$L_{\odot}$ for \mma\ and \mmb\ respectively. To compensate for this source of uncertainty each SDC335 point on Figure \ref{Fout_Lbol:fig} is presented with a $\pm$50\% error bar on the  $L_{bol}$ value. For outflows A and B the dominant source of error is the range of possible inclination angles, as such for these sources the errorbar represents the maximum and minimum values for $F_{CO}$ using the respective maximum and minimum possible inclination angles. For outflow C the $F_{CO}$ error includes the measured error on $F_{CO}$, prior to inclination angle correction due to noise within the data, plus a $\pm$5 degree error on the inclination angle used (57.3\degs for the average angle for randomly orientated outflows).

We can see from Figure \ref{Fout_Lbol:fig} that all outflows in SDC335 fit the general trend of outflow momentum flux as a function of luminosity over the wide mass range represented. The plotted points are the sum of the red plus blue lobes for each source in SDC335, using the temperatures from our RADEX fitting and derived optical depth values in the calculation of the $F_{CO}$ values. The outflows A and B in SDC335 reside in the same parameter space as the high-mass protostars from \citet{Maud15} with outflows A amongst the higher $F_{CO}$ range for its calculated $L_{bol}$. We note that the sources from \citet{Maud15} are studied with a single dish instrument, thus lacking the resolution of our current work, leading to the question posed by those authors regarding whether or not their observed outflows are driven by a single or multiple objects. SDC335 outflow C has a significantly lower $F_{CO}$ value away from the proposed observed trend seen within the literature. This suggests either it has a less powerful, potentially more evolved outflow or this effect could simply be an artefact caused by the limited velocity range we are able to integrate over for this outflow coupled with a poorly defined inclination angle. 

Assuming each of the identified outflows in SDC335 represents a single protostellar object, we can consider whether these sources represent an analogue of the protostellar source evolution classifications used at lower mass \citep{Lada99}. To this end, we generate best linear fit lines for the $F_{CO}$ as a function of $L_{bol}$ for Class 0 and I sources using values from the literature plotted in Figure \ref{Fout_Lbol:fig}. The best fit lines are plotted as the red dotted line for Class 0 and blue dashed-dot line for Class I, extended to higher luminosity values for comparison with SDC335 with associated 1-$\sigma$ error margins from the fit plotted as shaded regions of matching colour. 

The derived best fits are,
\begin{equation}
\log_{10}(F_{CO}[\rm{Class 0}])=-4.3(\pm0.15) + 0.60(\pm0.08)\log_{10}\left(\frac{L_{bol}}{L_{\odot}}\right)
\end{equation}
and 
\begin{equation}
\log_{10}(F_{CO}[\rm{Class I}])=-5.3(\pm0.12) + 0.63(\pm0.16)\log_{10}\left(\frac{L_{bol}}{L_{\odot}}\right)
\end{equation}

\noindent for Class 0 and 1 sources, respectively. We note that the best-fit line for Class 0 objects matches very closely to the results reported by \citet{CabritBertout92} for a sample of Class 0 sources (IRAS 16293, IRAS 3282, L1448, L1455M, RNO 43 and VLA1623) and a single high-mass object (G35.2 N).
\vspace{0.5cm}

The fitted lines show a decrease in outflow momentum flux between the Class 0 and I stages, as would be expected during the evolution of a protostellar source along the plotted evolutionary tracks \citep{Ana13} assuming a decreasing  and intermittent accretion rate. From Figure \ref{Fout_Lbol:fig} we see that both \mma\ and \mmb\ lie very close to the Class 0 best fit line and within the 1-$\sigma$ error bound of that line of best fit (red shaded region). This suggests that these sources are very high-mass Class 0 analogue. We use the qualifier `very' to indicate 'more extreme than' based on a comparison to the position of the other high-mass Class 0 protostars in Figure \ref{Fout_Lbol:fig} from \citet{Ana13} (empty star markers in the Figure).

Supporting the `young, very high-mass' status of the SDC335 sources \mma\ and \mmb,\, we can see from the evolutionary tracks in Figure \ref{Fout_Lbol:fig} that these sources sit well above the 50\solmass\ track, but at an early stage before $\sim$50\% of the envelope mass has been accreted (denoted by the first arrow head). This is also corroborated by the short dynamical times of these outflows ($\rm{few}\times 10^3$yr) in Table \ref{outflowprop:tab}.

\citetalias{Avison15} found, based on their current radio continuum properties, that the three high-mass protostellar objects in SDC335 are all currently displaying characteristics of zero age main sequence (ZAMS) stars of spectral type B1.5 (or B1.5-B1 for \mma), which equates to stellar masses of $\sim$ 9.0\solmass\ \citep{Mottram11}. These relatively low `current' stellar masses agree with the position of the observed outflow properties for A and B in terms of their position along the evolutionary tracks.

%--- START CO OUTFLOW PROPERITES TABLE -----------
\begin{table*}
\caption[]{CO(3-2) Outflow properties derived from ALMA data. Values for $P_{CO}$, $E_{CO}$ and $F_{CO}$ are corrected for the respective inclination angles given in Table \ref{outflowprop:tab}. We use the average value of the correction factor function over the range of angles for outflows A and B (see Tables \ref{outflowprop:tab} and \ref{CorrectionFactors:tab}) and the single angle value for C. Two sets of values are given for each outflow lobe, the first at T=20K using a literature $\tau_{corr}$ factor of 3.5, the second using our derived temperature and $\tau_{corr}$ factor (8.2) values. See text for more information. } %{\color{red}UPDATE ALL THIS!!!}}
\begin{center}
\begin{tabular}{c c c c c c c c}
\hline
\hline
Outflow & Lobe & T  & $M_{out}$ & $P_{CO}$ & $E_{CO}$ & $F_{CO}$ & $\dot M_{acc}$\\
 & & [K] &  [\solmass] & [\solmass \kms] &  [$10^{38}$J] & [10$^{-5}$\solmass \kms yr$^{-1}$] & [10$^{-5}$\solmass yr$^{-1}$] \\
\hline
\multirow{4}{*}{A} & Blue & 20 & 0.19 &  8.72 &  4.51 &  274.1 & 11.0\\
 &  Red & 20  & 0.89 & 48.0 &  29.2 &  1619.1 & 64. 8\\ 
 & Blue & 62.0 & 0.75 & 12.7 & 6.5 & 368.6 & 14.7\\ 
 & Red & 53.3  & 1.27 & 68.2  & 41.4 & 2133.8 & 85.4 \\ 
\hline
\multirow{4}{*}{B} & Blue & 20 & 0.08 & 4.6 &  2.98 & 218.6 & 8.7\\ 
 & Red & 20 & 0.21 &  13.2 &  9.6 & 461.0 & 18.4 \\ 
 & Blue & 54.6 & 0.22 & 12.1 & 7.5 & 288.0 & 11.5 \\ 
 & Red & 60.9 & 0.32 & 20.0  & 14.4 & 617.9 & 24.7\\ 
 \hline
\multirow{2}{*}{C} & Blue & 20  & 0.002 & 0.07 & 0.02 & 6.1 & 0.2\\ 
 & Blue & 54.5  & 0.014 & 0.31 & 0.07 & 8.0 & 0.3\\ 
\hline
\end{tabular}
\end{center}
\label{CO_outflowprop:tab}
\end{table*}
%--- END CO OUTFLOW PROPERITES TABLE -----------

\subsubsection{Low bolometric luminosities in the MM1 core}
\label{tooLow:sec}
An open issue from previous work on SDC335, presented in \citetalias{Avison15}, was the discrepancy between the observed bolometric luminosity $L_{bol}$, as derived from the millimetre core spectral energy distributions, and the luminosity of a ZAMS star of the spectral type necessary to produce the Lyman $\alpha$ photon flux inferred from the radio continuum, $L_{\rm ZAMS}$. The value of  $L_{bol}$  was found to be approximately a factor 20 less than $L_{\rm ZAMS}$. 

To briefly review the \citetalias{Avison15} finding (for more details, see their Section 4.1); the authors calculated a Lyman $\alpha$ photon flux based upon the HC\hii\ continuum flux density for each of the three HC\hii\ sources. From this, a ZAMS spectral type was associated with each core (B1.5 for MM2 and \mmb\ and B1 $-$ B1.5 for \mma) using values from \citet{Mottram11} and \citet{Davies11}. 

The ZAMS is reached when hydrogen burning has commenced, however, given the outflow indicators (masers, EGO etc) present in SDC335, it was assumed in \citetalias{Avison15} that each HC\hii\ region represented a protostar which was still actively accreting. This assumption is borne out by the detection of the three outflows in the current work. An actively accreting protostar will have a total luminosity (assuming ZAMS properties), $L_{tot,ZAMS}$, which is the sum of its intrinsic luminosity, $L_*$, and that from accretion, $L_{acc}$. Here $L_{acc}$ is of the form $L_{acc} = \frac{GM_*\dot M_*}{R_*}$, and the value of $M_*$,  $R_*$ and  $L_*$ were based on the ZAMS properties and $\dot M_*$ (the mass accretion rate) was assumed to be equal to the global infall rate of SDC335 derived by \citet{Peretto13} of $2.5\times10^{-3}$\solmass\ yr$^{-1}$. 

From this calculation, the authors noted that the ZAMS $L_*$ and $L_{bol}$ agree to within a factor of $\sim2$ however this did not allow for ongoing accretion. Including $L_{acc}$ gives $L_{tot,ZAMS}$ a value that is a factor 20 higher than $L_{bol}$ (see the values in Tables 2 and 6 of \citetalias{Avison15}). 

\citetalias{Avison15} presented two scenarios which could account for the observed low bolometric luminosity from all three protostars in SDC335. In light of the data presented in this current work we can review these scenarios. The first scenario was that the assumed accretion rate $2.5\times10^{-3}$\solmass\ yr$^{-1}$ from \citet{Peretto13}) was too high and that the protostars were undergoing accretion at a lower rate (either overall or as a function of periodic accretion). In this current work, we derive a mass accretion rate onto each protostellar object from their outflow momentum flux (column 8 in Table \ref{CO_outflowprop:tab}) and find that indeed the values are lower than the \citet{Peretto13} infall rate. This allows us to revise the values of $L_{tot,ZAMS}$ from Table 6 of \citetalias{Avison15} and address the second scenario discussed by those authors. In doing so, using both the minimum and maximum $\dot M_{acc}$ for each outflow from Table \ref{CO_outflowprop:tab} to give a possible range of $\dot M_{acc}$ (and using the derived temperatures from \S\ref{Radex:sec}), we provide revised values for Table 6 in \citetalias{Avison15} in our current Table \ref{RevisedZams:tab}. 

%----------------------------- START AVison 15 TABLE 6 revised -----------------------------
\begin{table*}
\caption{Calculated luminosity values for the three protostellar cores in SDC335, assuming the ZAMS properties associated with each source in \citetalias{Avison15}. This is a revised and expanded version of Table 6 in \citetalias{Avison15}. For each source, we find a minimum and maximum $L_{acc}$ and this $L_{tot,ZAMS}$ based upon the range of $\dot M_{acc}$ between the red and blue outflow lobes from Table 5. $^{\dagger}$ $L_{bol}$ as derived in \citetalias{Avison15}, the division of $L_{bol}$ for the MM1 core is described within $\S$\ref{EvoStat:sec} of this paper.}
\begin{center}
\begin{tabular}{c c c c c c c c}
\hline
\hline
 & $L_{*}$ & $L_{bol}^{\dagger}$ & $L_{acc}$ (min) & $L_{acc}$ (max) & $L_{tot,ZAMS}$ (min) & $L_{tot,ZAMS}$ (max) & $ L_{tot,ZAMS}/L_{bol}$ factor \\
 Source & [L$_{\odot}$]  & [L$_{\odot}$]  & [L$_{\odot}$]  & [L$_{\odot}$]  & [L$_{\odot}$] & \\
\hline
\mma\ & 5.5$\times 10^{3}$ & 1.3$\times 10^{4}$ & 1.3$\times 10^{4}$ & 7.6$\times 10^{4}$ & 1.9$\times 10^{4}$ & 8.1$\times 10^{4}$ & 1.4 - 6.2\\
\mmb\ & 4.1$\times 10^{3}$ & 5.1 $\times 10^{3} $ &9.9$\times 10^{4}$ & 2.1$\times 10^{4}$ & 1.4$\times 10^{4}$ & 2.6$\times 10^{4}$ &  2.8 - 5.0\\
MM2      & 4.4$\times 10^{3}$ & 9.9$\times 10^{3}$ & 1.7$\times 10^{2}$ & $-$ & 4.5$\times 10^{3}$ & $-$ & 0.5\\
\hline
\end{tabular}
\end{center}
\label{RevisedZams:tab}
\end{table*}
%----------------------------------- END AVison 15 TABLE 6 revised ---------------------------------

For MM2, the $L_{tot,ZAMS}$ is now a factor $\sim$2 lower than the $L_{bol}$ which is consistent within the uncertainties on these values. A spectral type B1.5 ZAMS star has a mass of $\sim$ 9\solmass\ \citep{Mottram11}, which for MM2 would also agree with the accreting protostars position on the evolutionary tracks in Figure \ref{Fout_Lbol:fig} and its status as a potentially more evolved protostar than the cores in MM1. 

In the case of the two ionising sources in MM1, the derived mass accretion rates cover a large range of values, 14.7 to 85.4$\times10^{-5}$ \solmass\ yr$^{-1}$ and 11.5 to 24.7$\times10^{-5}$ \solmass\ yr$^{-1}$ for  \mma\ and \mmb, respectively. With these values the bolometric luminosity for \mma\ is consistent with $L_{tot,ZAMS}$ at the lowest $\dot M_{acc}$ values but the discrepancy persists over the majority of the possible range (any value above 30.0 $\times10^{-5}$ \solmass\ yr$^{-1}$ leads to a factor 2.5 discrepancy between the two luminosities). And for \mmb\ the discrepancy between $L_{bol}$ and $L_{tot,ZAMS}$ is at a factor of about three over the entire range -- meaning that the observed $L_{bol} < L_{tot,ZAMS}$ from \citetalias{Avison15} also persists for this source.

The second scenario to account for the
luminosity discrepancy discussed in \citetalias{Avison15} is based upon the models of massive protostellar evolution at high accretion rates ($\sim 10^{-3}$\solmass\ yr$^{-1}$) of \citet{HosokawaOmukai09} and \citet{HosokawaYorkeOmukai10}. In this model massive protostars go through a phase of swelling to radii of $\sim100$ R$_{\odot}$ when their mass is between 6 and 10\solmass, followed by a short contraction phase as their mass grows to between 10 and 30\solmass. During the period of swollen radii the effective temperature of the protostar is lower than that of an equivalent mass ZAMS star, as such there are insufficient UV photons to generate a \hii\ region. However, during the short contraction phase as the radius decreases the effective temperature increases and a \hii\ region can form, whilst the radius remains somewhat swollen ($\sim$10 R$_{\odot}$) giving a lower luminosity than a ZAMS star at the same mass. In \citetalias{Avison15}, the authors suggest that each SDC335 protostar was in the contraction phase, accounting for both the observed \hii\ regions and the low bolometric luminosities, though given the relatively brief duration of the contraction phase having multiple sources at this stage simultaneously would be unlikely. 

The new derived mass accretion rates for \mma\ and \mmb\ have values that are in the range of those considered by the \citet{HosokawaOmukai09} models (although for the swollen radii at accretion rates of $\sim 10^{-4}$\solmass\ yr$^{-1}$, the swelling is to $\sim$40R$_{\odot}$ rather than 100R$_{\odot}$) and thus these sources are in either the swollen radii phase or the contraction phase -- this would still seem a viable scenario that would lead to the lower observed luminosities. Given the relative lengths of each of these phases, it would be more likely that they are still in the swollen radii phase. If this is the case, the origin of the ionised emission in each core would then require an alternate explanation, which we address in Section \ref{radiojets:sec}.

%--- OUTFLOW SHOCKS
\subsubsection{An alternate interpretation of the ionised radio continuum emission in SDC335}
\label{radiojets:sec}
As discussed briefly in \citetalias{Avison15}, the spectral indices for \mmb\ and MM2 fall within the range of values for both photo-ionised \hii\ regions and collimated ionised jets \citep{Reynolds86}. Given the detection in this work of outflows from both these sources, it is important to review origin of radio continuum emission from the SDC335 protostars.

Radio continuum emission has been detected toward a number of low luminosity sources driving molecular outflows, for example in \citet{Anglada95}. In this work, the author shows that the low luminosity for their sample of objects precludes the origin of the observed radio continuum emission coming from photoionisation by Lyman alpha photons. Using the models of shock ionised gas from \citet{Curieletal87, Curiel89}, the author then shows that for their sample, shock ionisation is capable of creating the observed radio continuum based on the outflow momentum flux of their targets.

Based upon the best-fit model and data in \citet{Anglada95} (their Equation 1 and Figure 5) we find that using their 8, 23, and 25~GHz flux densities \citetalias{Avison15},  \mma\ and \mmb\ are shown to reside between the line of minimum requirement (dashed line in \citet{Anglada95} Figure 5) and the line of best fit (solid line in \citet{Anglada95} Figure 5) for shock ionisation as the mechanism for the observed radio continuum emission in their sample. This suggests that for these sources this may indeed be the origin of the detected radio emission. However, MM2  is  a factor of between $\sim$17 and 37 below the lower bound of the emission expected from shocks \citep{Anglada95}. This may be an evolutionary factor or one arising from the orientation of the MM2 outflow to the line of sight, which limits the range of velocities we are able to integrate over leading to an underestimate of the outflow momentum flux value for this source (\S \ref{infall:sec}).

The direct implication of interpreting the origin of the ionised matter as from outflow shocks as opposed to photoionisation is the evolutionary status of the protostellar sources themselves. If the emission is not due to photoionisation, this suggests that the protostars are at an earlier pre-ionising stage in their evolution. 

It is important to state that although this alternate interpretation may account for the presence of ionised hydrogen and indicate a resolution of the discrepancies between observed bolometric luminosity and a corresponding ZAMS luminosity, the protostars within SDC335 remain high-mass protostellar sources, which will go on to form stars of mass $>8$\solmass. We base this on both their observed luminosities (of the order 10$^4$ L$_{\odot}$ \citetalias{Avison15}) and the co-location of each of the radio emission peaks with 6.7GHz methanol maser emission. Higher spatial resolution ($\leq$ 2.0 \asec) observations of SDC335 at radio frequencies would be required to fully assess this interpretation. The impact on the current work -- should the radio free-free emission prove to be outflow shocks -- is limited to the discussion in Section  \ref{tooLow:sec}. All other results would remain unchanged.

%-------------------------------------------------------------------------------------------------------------------
%                             Cloud Outflow interaction 
%-------------------------------------------------------------------------------------------------------------------
\section{Interaction between outflows and filaments }

Previous studies of the motion of material at large scales in SDC335 have found that the cloud is both globally collapsing toward its centre and that material is being transported inward to this region along the filamentary arms of the cloud \citep{Peretto13}.  In our current work, however, we report the detection of material outflowing from the protostellar cores at the cloud centre. In this section, we look at the evidence of the interaction of the filamentary infalling and outflowing material. Finding regions of interaction within SDC335 would make the IRDC a valuable source for studying the potential disruptive effects on material transport from feedback of young massive stars have.
In the following, we highlight the observed features within SDC335 that mark potential filament-outflow interactions. As part of this discussion, we identify each outflow lobe by the letter corresponding to the outflow and a sub-script colour to indicate the particular red or blue lobe.

%---------------------------------------
%  MASERS
%---------------------------------------
\subsection{Class I methanol masers}
\begin{figure*}
\centering
\includegraphics[scale=0.35]{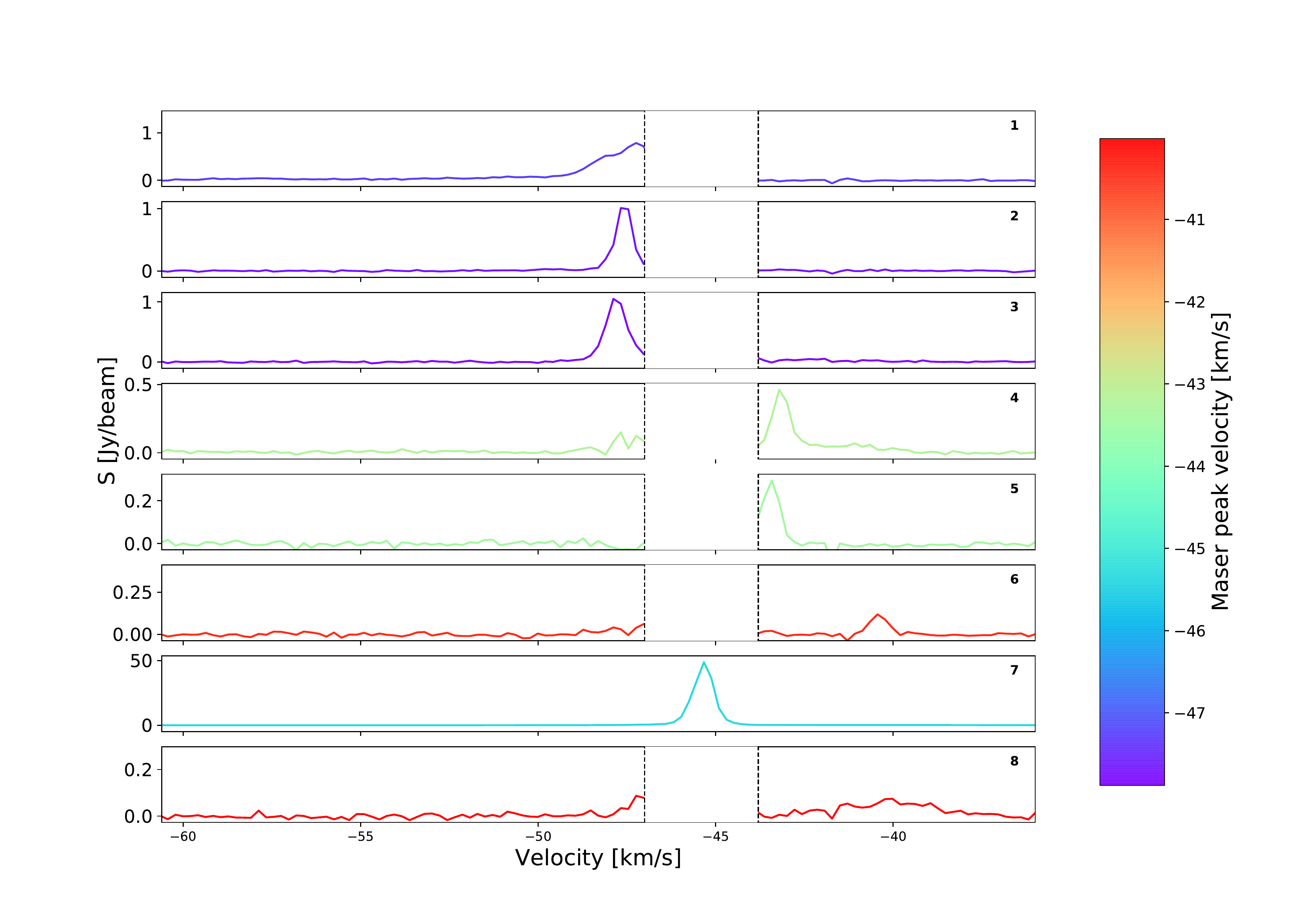}
\caption{Spectra taken at the peak position of the eight Class I methanol masers observed toward SDC335. Each panel gives the spectra of an individual maser spot (numbered in the upper right of the panel). Each spectra is colour coded by its peak velocity following the colour bar (\textit{right}).  The $y$-axis for each spectrum differs and is scaled to between -0.1$\times$ to 1.1$\times$ the peak flux density of the individual maser as listed in Table \ref{Maser:tab}. The velocity range $-47.0$ to $-43.8$\kms\ is blanked out in panels 1 - 6 and 8 as between these velocities the spectra of these masers are dominated by imaging artefacts caused by the strong emission from maser 7.  This is to avoid confusion between true maser emission and imaging artefacts.}
\label{MaserSpec:fig}%
\end{figure*}

\begin{figure}
\centering
\includegraphics[scale=0.35]{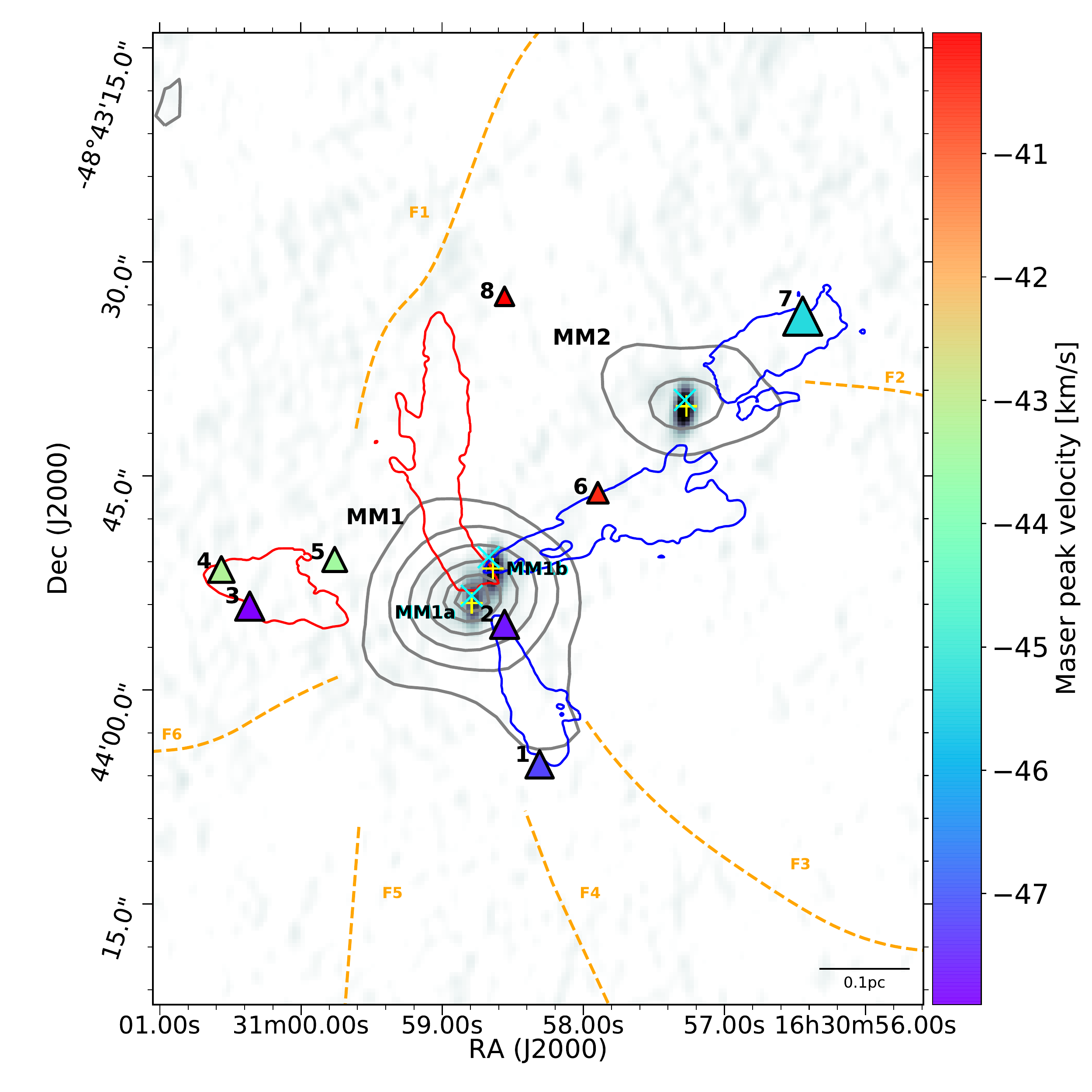}
\caption{SDC335 region showing the position of the Class I methanol masers. Orange dashed lines as per Fig. \ref{PressyFig:fig}. Filled triangles, the Class I methanol masers detected in the ATCA data, colour-coded by velocity matching that in Fig. \ref{MaserSpec:fig}, the size of the triangle is proportional to the flux density of the maser source. Cyan `$\times$' denote the location of Class II methanol masers at 6.7GHz and yellow `$+$' denote the \water\ masers at 22GHz. Grey contours show the ALMA 3mm continuum emission and the greyscale images shows the ATCA 8GHz continuum emission. The red and blue contours show $\sim$9\% of the peak integrated intensity of each of outflow, using the velocity ranges from Table \ref{OutflowCalc:tab}  as a guide when comparing the maser positions.  }
\label{MaserCO:fig}% \LEt{ Please denote Class II or Class 2 consistently within the paper, also maintaining the uppercase C.}
\end{figure}

Collisionally excited Class I methanol masers are commonly associated with molecular outflows in regions of massive star formation \citep[e.g.][]{Kurtz04, Cyganowski09}. SDC335 was known to harbour four Class I maser sources \citepalias[and references therein]{Avison15}, all of which are clearly spatially offset from the compact \hii\ regions in the MM1 and MM2 cores. With our new, more sensitive data, an additional four individual maser sources were detected. The spectrum of each maser is presented in Figure \ref{MaserSpec:fig}. Figure \ref{MaserCO:fig} presents the maser locations within SDC335, with the masers colour-coded by velocity.

All the maser spots peak at velocities within $\sim$6.0\kms\ of the systemic velocity of the SDC335 mm-cores. The position and peak emission properties of each spot are given in Table \ref{Maser:tab}. The masers are numbered from south to north. Whilst we do not focus on the properties of each maser spot individually within this work, we include in the following discussion the maser spots that provide useful indications as shock tracing when interpreted as potentially part of the outflow-filament interactions within SDC335.

%----------------------------- START Maser Properties TABLE -----------------------------
\begin{table*}
\caption[]{Characteristics of observed Class I \meth\ masers. Maser spots within the table are listed and indexed from south to north. Masers 1 - 5 have velocity profiles which extend into the same velocity range of maser 7 (denoted by the left or right arrows), where it is not possible to correctly identify velocity features of the source or sidelobe artefacts of maser 7. }
\begin{center}
\begin{tabular}{c c c c c c}
\hline
\hline
Maser & RA & Dec & S$_{peak}$ & V$_{peak}$ & V$_{range}$ \\
No. & {[h : m : s]} & {[ $^{\circ}$ : $^{\prime}$ : \asec]} & [Jy] & [\kms] & [\kms]\\
\hline
1 & 16:30:58.31 & -48:44:05.2 & 0.79 & -47.3 & -60.4, -47.0$\rightarrow$\\
2 & 16:30:58.56 & -48:43:55.4 & 1.01 & -47.7 & -52.3, -47.0$\rightarrow$\\
3 & 16:31:00.36 & -48:43:54.1 & 1.05 & -47.9 & -49.6, -47.0$\rightarrow$\\
4 & 16:31:00.56 & -48:43:51.6 & 0.462 & -43.2 & $\leftarrow$-43.8, -39.0\\
5 & 16:30:59.76 & -48:43:50.9 & 0.294 & -43.4 & $\leftarrow$-43.8, -42.3\\
6 & 16:30:57.90 & -48:43:46.2 & 0.119 & -40.4 & -41.5, -39.2\\
7 & 16:30:56.45 & -48:43:33.8 & 48.8 & -45.3 & -52.4, -39.0\\
8 & 16:30:58.56 & -48:43:32.4 & 0.0742 & -40.0 & -43.0, -36.8\\
\hline
\end{tabular}
\end{center}
\label{Maser:tab}
\end{table*}
%----------------------------- END Maser Properties TABLE -----------------------------

%------------------------------ START Maser Spectra ------------------------------------------
%------------------------------- END Maser Spectra --------------------------------------------

%------------------------------ START Maser Position ------------------------------------------
%------------------------------- END Maser Position --------------------------------------------

%---------------------------------------
% Interaction discussion
%---------------------------------------
\subsection{Interaction regions}
\subsubsection{A$_{Blue}$ and the F3 and F4 filaments}
\begin{figure*}
\centering
\includegraphics[scale=0.7]{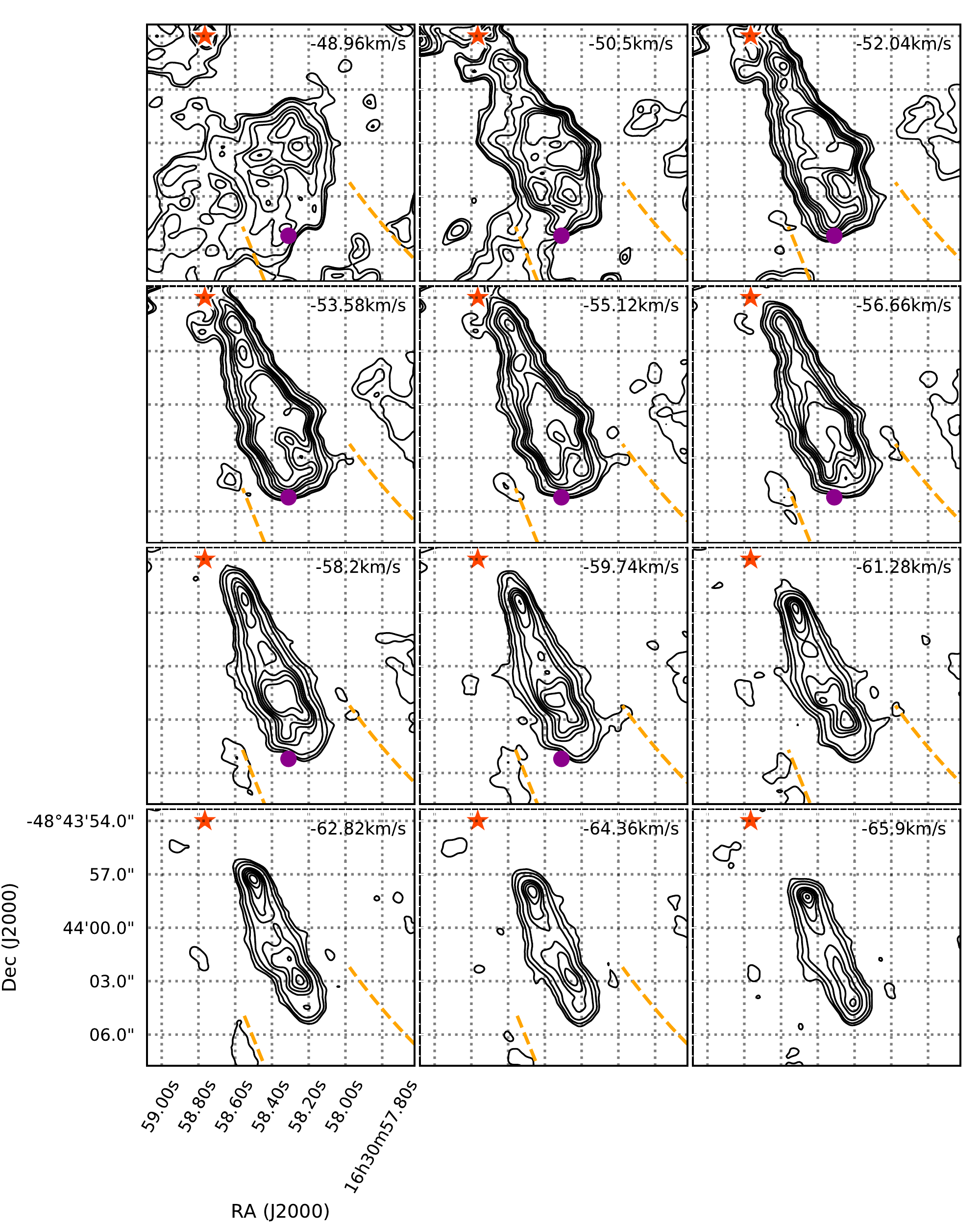}
\caption{Channel map of the CO emission from the A$_{Blue}$ outflow. The velocity of each image is listed in the upper right corner. The position of MM1a is indicated by the red star. The purple filled circle gives the position of the Class I \meth\ maser source 1 (see Table \ref{Maser:tab}), only at velocities where the maser is present.  The orange dashed lines show the fiducial directions of the F3 and F4 filaments as per Figure \ref{PressyFig:fig}. Contours are plotted at the 5,10, 20, 40, 50, 60, 70 and 80 $\sigma$ level of the ALMA CO image (where $\sigma$ = 22.0 mJy/bm).}
\label{BlueJetVcutCO:fig}%
\end{figure*}

The A$_{Blue}$ outflow is well collimated near the driving source (\mma) when observed at high resolution (ALMA CO and \thirCO), as seen in the narrow structure of the contour plots in Figure \ref{OUTFLOW_GRID:fig}-A$_{zoom}$ and the CO velocity channel maps shown in Figure \ref{BlueJetVcutCO:fig}. The shock-tracing SiO emission detected in this outflow lobe is however predominately at the end (furthest from the driving source) of this lobe Figure \ref{OUTFLOW_GRID:fig}-B.

This region of SiO emission and the observed lobe end in the CO and \thirCO\ ALMA observation coincide spatially and kinematically with the ends of the F3 and F4 filaments of the IRDC (marked as the orange dash line in Figures \ref{PressyFig:fig} and \ref{OUTFLOW_GRID:fig}, c.f. \citealt{Peretto13}). The F3 and F4 filaments are carrying material into the SDC335 central region at velocities of $\sim-$45.8 and $-46.5$\kms\, respectively \citep[see][their Fig. 4c and 5]{Peretto13}. Given the $V_{lsr}$ of the \mma\ core, these infalling velocities are moving in the opposite direction to the outflowing material ($V_{Ablue} \sim$  $-$50.5 to $-$82.4\kms), see Figure \ref{BlueJetVcutCO:fig} and \ref{OutflowCalc:tab}). Supporting this interpretation is the position and velocity range covered by the Class-I \meth\ maser source 1 (see Table \ref{Maser:tab}). We show the presence of the maser spot at a given velocity in Figure \ref{BlueJetVcutCO:fig} as a solid purple circle. 

Given the collisionally excited pumping mechanism responsible for Class-I masers and the hypothesis that such collisional excitement occurs at the interfaces of outflows and the surrounding molecular gas as seen in, for example, \citet[][]{Plambeck90, Kurtz04, Vornokov10}, the alignment of the maser activities both spatially and kinematically with the outflow lobe end suggest that we are indeed seeing the point of interaction between the A$_{Blue}$ outflow and the infalling material from the F3 and F4 filaments. The large velocity range the \meth\ emission covers requires a strong shock front to pump such maser emission.

%------------------------------ START Blue Jet Blob CO ------------------------------------------
%------------------------------- END Blue Jet Blob CO --------------------------------------------

\subsubsection{C and the F2 filament}

The one detected lobe of outflow C (C$_{blue}$) exhibits both the brightest Class I methanol maser (maser 7, Table \ref{Maser:tab}) and the brightest SiO emission detected in the region (see Figure  \ref{OUTC_SPEC:fig}). At a peak flux density of 48.78\,Jy, maser 7 is over 40 times the intensity of any of the other detected maser sources (see e.g. Figure \ref{MaserSpec:fig}). Given that both SiO emission and Class I \meth\ maser are shock tracing species we note that the end (away from the driving source MM2) of this outflow lobe (Figure \ref{OUTFLOW_GRID:fig}) appears close to the fiducial end of the F2 filament (orange dashed line, c.f. \citealt{Peretto13}). This suggests that we are observing a shocked region caused by the meeting of the outflow and the inflowing matter from the F2 filament. The velocities of C$_{blue}$ and the infall along F2 are in opposite directions, with the F2 filament $V_{F2}=\sim-$46.0\kms\ \citep[see][their Fig. 4c and 5]{Peretto13}, so redder than the $V_{lsr}$ of MM2 and $V_{C_{blue}}$=$-$59.1 to -43.6\kms,\ thereby extending both red and blue-ward of this $V_{lsr}$. Although the maser position and the end of the orange dashed line representing F2 in Figures \ref{OUTFLOW_GRID:fig}-A,B,C do not match exactly, we emphasise that the dashed line is a fiducial marker and the true morphology and kinematics at the end of the inflow are not known at spatial resolutions better than $\sim$5\asec \citep{Peretto13}. Further higher resolution observations of the region would be required to confirm this as a true shock-interaction region. 

\subsubsection{Curvature of A$_{red}$ and the F1 filament}
\label{CurveRed:sec}
\begin{figure}
\centering
\includegraphics[scale=0.55]{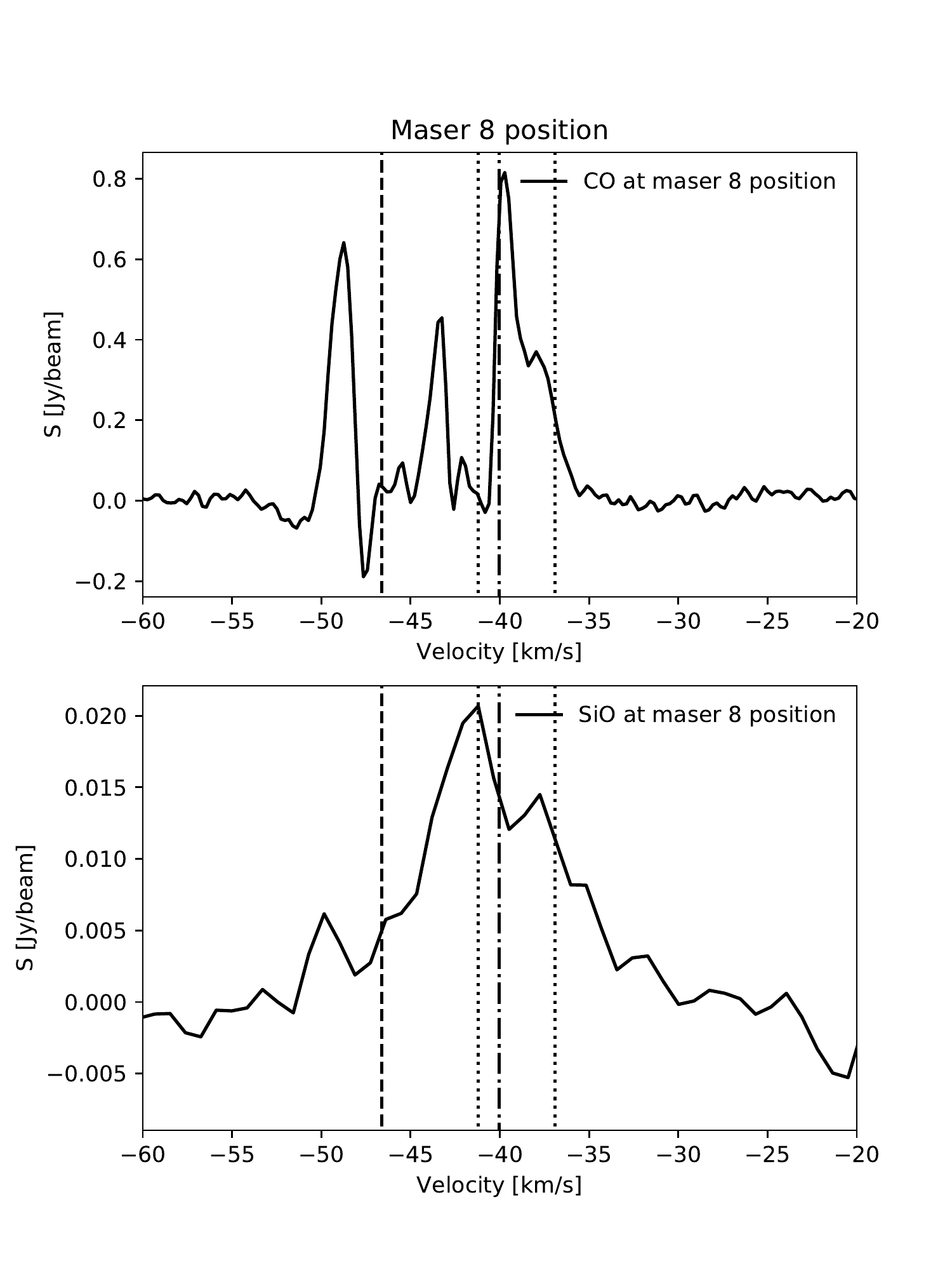}
\caption{CO and SiO spectra measured at the position of the maser 8 peak emission see Table \ref{Maser:tab}.  The vertical dashed line gives the $V_{lsr}$ of the target and the vertical dotted lines mark the range in velocity of the clump of emission seen toward the curvature of the A$_{red}$ outflow lobe discussed in \S \ref{CurveRed:sec}. The vertical dash-dot line gives the velocity of peak emission from maser 8.}
\label{maser8_COSiO:fig}%
\end{figure}

The A$_{red}$ lobe is observed in our CO (ALMA), SiO, and HNC data to exhibit curvature at its northernmost end. Inspecting the position of the nearby filamentary arm, F1, in Figure \ref{OUTFLOW_GRID:fig} it appears that the curvature coincides with where A$_{red}$ would meet the infalling material from the filament. Emission from the A$_{red}$ here is at velocities of $\sim -$30 to $-$20 \kms\ (CO, SiO) whereas the inflowing material of the F1 filament is at $\sim-$47.25\kms\ \citep[see][their Fig. 4c and 5]{Peretto13}. The start of curvature of the F1 filament also coincides with a clump of SiO emission and CO (beyond the velocity range used to generate the contours in Figure \ref{OUTFLOW_GRID:fig}-A) offset from the outflow (see Figure \ref{OUTFLOW_GRID:fig}-B) at velocities $-36.9$ to $-41.2$\kms. This  is spatially and kinematically coincident with maser 8 (see Table \ref{Maser:tab}), suggesting an interaction point between the inflow and outflow is likely associated with a shock front leading to the SiO. Maser 8 peaks at $-$40\kms\ with a velocity range $\pm\sim3.0$\kms\ as such covers velocities which are between those measured for the inflowing and outflowing material. Figure \ref{maser8_COSiO:fig} gives the spectra of CO and SiO at the position of maser 8.

%---------------------------- START Maser 8 SPECTRA -----------------------------------
%---------------------------- END Maser 8 SPECTRA -----------------------------------

\subsubsection{B$_{red}$ shocked emission}

\begin{figure}
\centering
\includegraphics[scale=0.55]{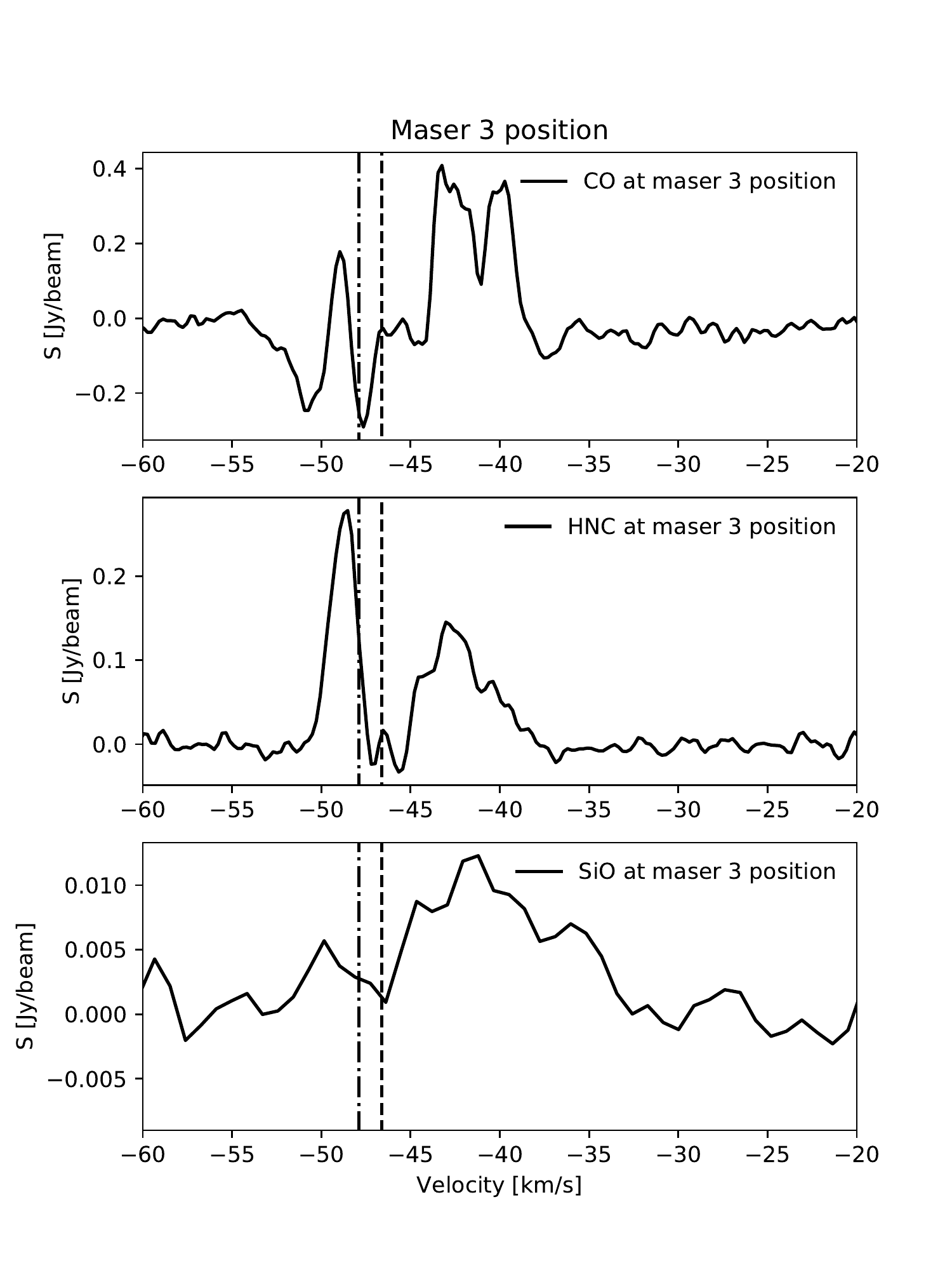}
\caption{CO (ALMA), HNC and SiO spectra measured at the position of the maser 3 peak emission see Table \ref{Maser:tab}.  The vertical dash-dot line gives the $V_{lsr}$ of the outflow driving source, \mmb.}
\label{maser3_COSiO:fig}%
\end{figure}

\begin{figure}
\centering
\includegraphics[scale=0.55]{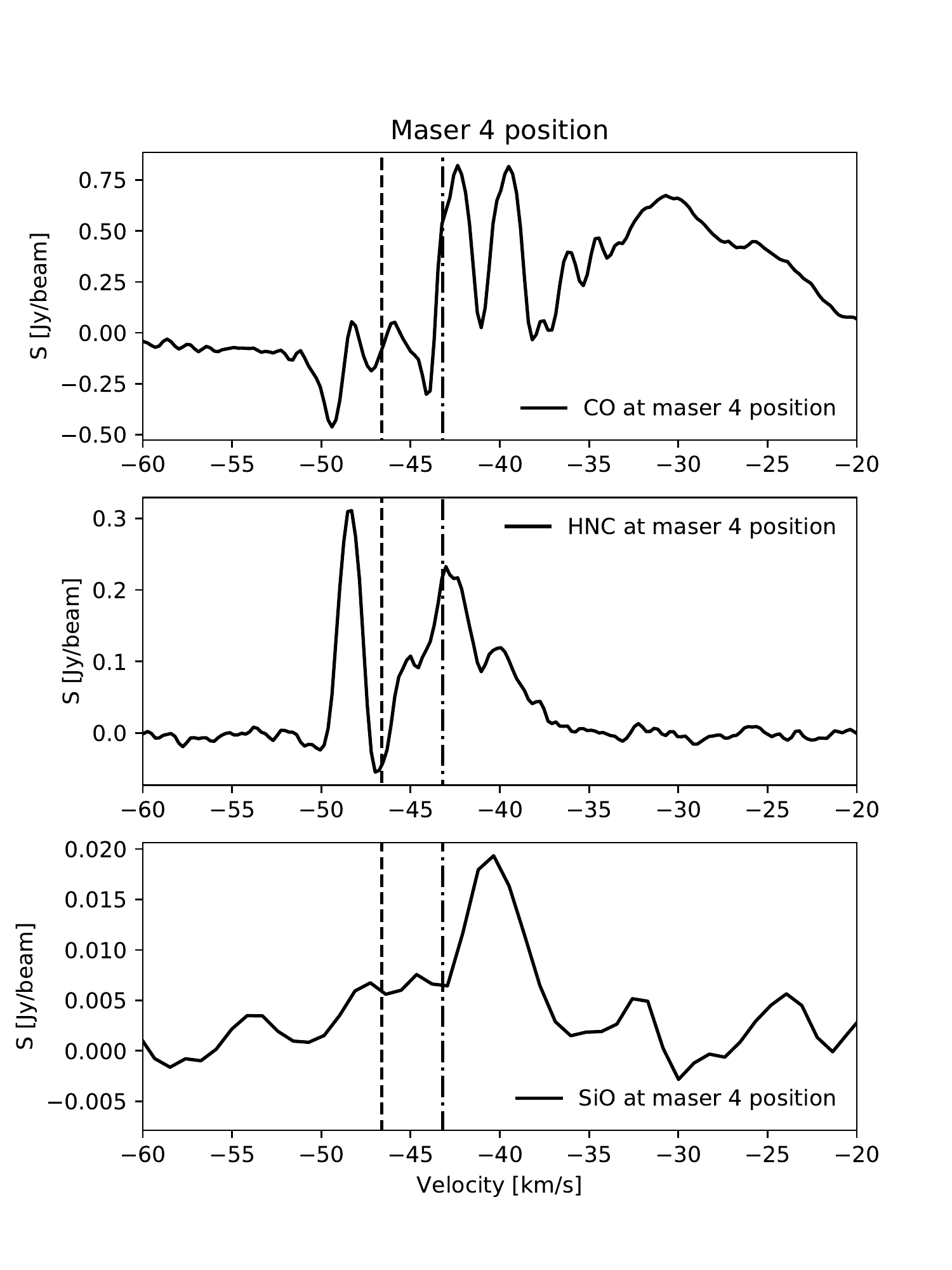}
\caption{CO and SiO spectra measured at the position of the maser 4 peak emission see Table \ref{Maser:tab}.  The vertical dash-dot line gives the $V_{lsr}$ of the outflow driving source, \mmb.}
\label{maser4_COSiO:fig}%
\end{figure}

The B$_{red}$ lobe is observed in SiO, HNC and CO (both APEX and ALMA). The leading edge of this outflow is coincident with the positions of masers 3 and 4 and peak SiO emission (see Figures \ref{OUTFLOW_GRID:fig}-A and \ref{OUTFLOW_GRID:fig}-B) of the outflow, indicating a region of shocked gas. We note that this lobe does not directly appear to be interacting with a filamentary inflow, but from Figure \ref{PressyFig:fig} the outflow appears toward the edge of the 8\mewm\ dark region of the cloud suggesting some interaction at the cloud boundary. In Figures \ref{maser3_COSiO:fig} and \ref{maser4_COSiO:fig} we give the spectra of CO (ALMA), HNC and SiO at the position of masers 3 and 4.

%---------------------------- START Maser 3,4 SPECTRA -----------------------------------

%---------------------------- END Maser 3,4 SPECTRA -----------------------------------

\subsection{A and B outflows misalignment, a potential scenario}
\label{RosenSims:sec}

As evident in Figure \ref{OUTFLOW_GRID:fig} and noted in Section \ref{DescB:sec}, outflows A and B have outflow axes that are approximately perpendicular to one another.  Naively, one would assume that the angular momentum axes for a pair of protostars of similar mass separated by a relatively small distance ($\sim$9000AU) would be aligned. This appears not to be the case for SDC335. 

Within the literature, there are other examples showing similar misalignments. For example, protostars A and B within IRAS16293 (separated by 600 AU) were found to have rotation axes misaligned \citep[and references therein]{Pineda16293} with A observed edge on and B either having little to no observable rotation or a rotation axis in the plane of the sky. This gives A and B effectively a $\sim$ 90\degs\ offset in their angular momentum axes. Similarly, there are several systems found in Perseus (with pairs separated by 1000 to 10,000 AU) which show outflow axes that are misaligned between low mass protostars with the distribution of axes appearing to prefer random or anti-aligned orientations \citep{Lee16}.

We consider a possible cause for this misalignment in SDC335. This scenario posits that the outflow progenitors \mma\ and \mmb, after fragmenting from the MM1 mm-core, have grown in mass by accreting matter from different material flows arriving at each core from different directions, thus changing their respective angular momentum vectors as the system evolves. The high, non-uniform accretion rates inherent to massive star formation ($\dot M_{\rm acc} \sim 10^{-3} \; \rm M_{\rm \odot} \, yr^{-1}$) will push the star around and cause the angular momentum vector to precess. Such behaviour has been seen in three-dimensional (3D) radiation-magnetohydrodynamic (RHD) simulations of the collapse of a turbulent massive pre-stellar core into a massive star \citep{Rosen16, Rosen20}. The movement of the star and its angular momentum vector will cause the outflows to be launched over a larger area as the star grows in mass. This effect will broaden the entrained outflows since the protostellar outflows are likely launched along the star's or surrounding accretion disk's angular momentum vector (e.g. \citealt{Shu00,Pudritz07}). 

\begin{figure*}
\centering
\includegraphics[scale=0.15]{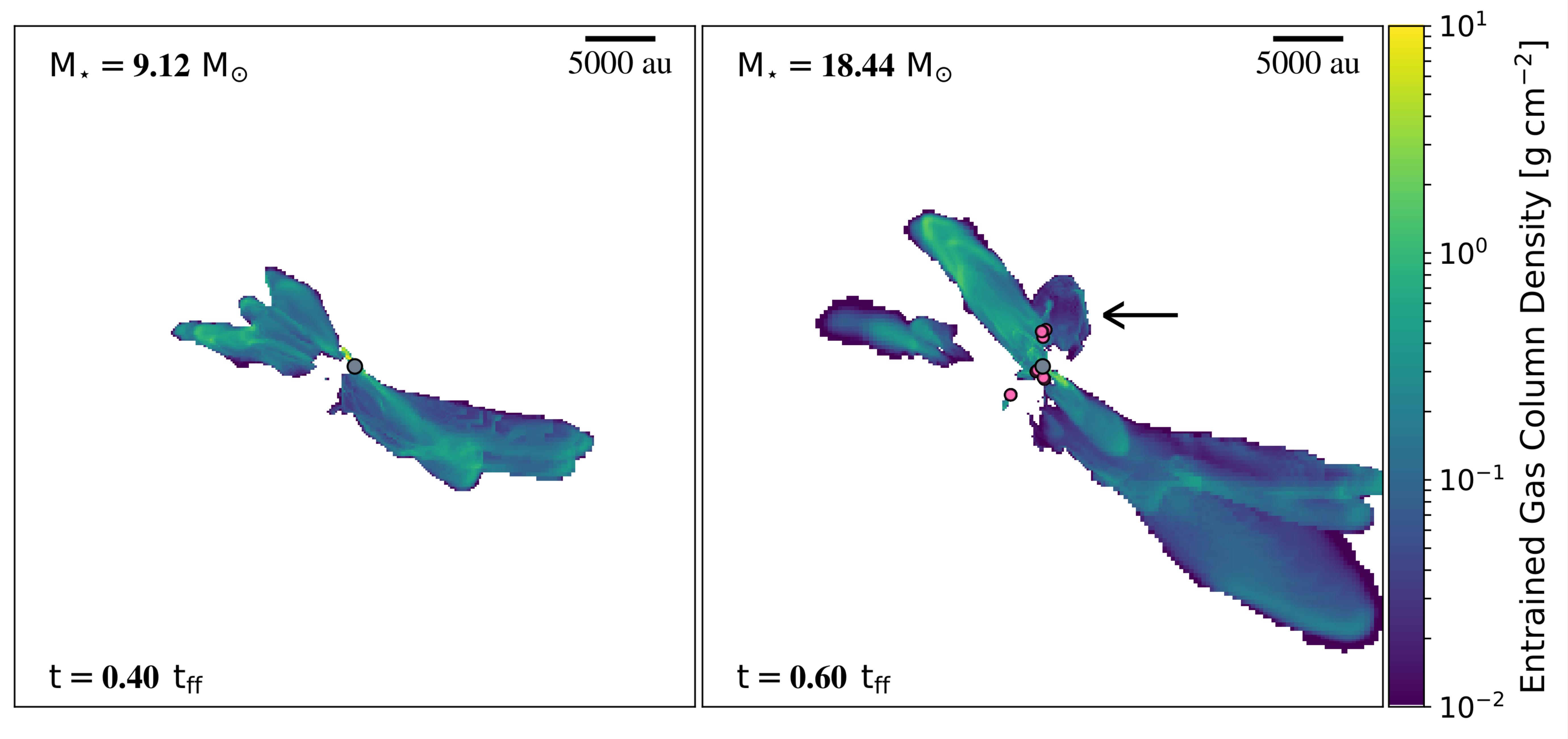}
\caption{Projection plots generated from simulations of high-mass star forming regions, \citep[adapted from][]{Rosen20}. These plots show only the material contained in the systems outflows in the condition $\frac{\rho_{outflow}}{\rho} > 0.1 $, at two time stamps during the free fall time of the star forming core. \textit{Left:} Solitary massive protostar (grey circle) is seen at $t=0.4 t_{ff}$. This massive protostar is at the centre of up to three `overlapping' outflows. These are thought to originate from the same outflow and their difference in position and angle is caused by precession of the star's spin axis and the movement of the protostar from its birth position to its current position within the simulation.  \textit{Right:} At later times, $t=0.6 t_{ff}$, we see that lower mass protostars have formed (pink circles) and are generating their own outflows. One such outflow, denoted by the arrow, is seen to be significantly offset in angle to the primary outflow from the massive star in the region.}
\label{Rosen:fig}%
\end{figure*}

Figure~\ref{Rosen:fig} \citep[adapted from][]{Rosen20} shows two snapshots from a 3D RHD simulation that models the collapse of a turbulent massive (150 $M_{\rm \odot}$) pre-stellar core into a massive stellar system and includes radiative feedback and collimated outflows that are launched along the stars' angular momentum vectors \citep[see subgrid model description by][]{Cunningham11}. These snapshots show the column density of the molecular material that is entrained by the protostellar outflows from the stars that have formed. To obtain the distribution of entrained material, we compute the mass-weighted integrated gas density only along cells that have $\rho_{\rm OF}/\rho \geq 0.1,$ where $\rho_{\rm OF}$ is the outflow gas density injected by stars, and $\rho$ is the total gas density.

In the first snapshot, where the time elapsed is 0.4 $t_{\rm ff}$ where $t_{\rm ff}=42.7$ kyr is the core's free-fall timescale, the star has a mass of 9.12 $M_{\rm \odot}$ and what looks like multiple molecular outflows. However, this outflow morphology is due to the non-uniform accretion flow that knocks the star around, causing the star to move away from its birth site and precess; thereby causing the outflow launching direction to change with time. At later times, $t=0.6 t_{ff}$, when multiple low-mass protostars have formed and are clustered near the massive protostar (pink filled-in circles), the smaller and weaker multiple outflows from the low mass protostars overlap. The resulting overlapping entrained outflows from these clustered low-mass companions to the right of the primary star are inclined by $\sim 45^{\circ}$  to the entrained outflow from the primary star, as indicated by the arrow in Figure~\ref{Rosen:fig}. We note that this significant offset between outflow axes and separation between the massive and low-mass stars is similar to what is seen in SDC335.
The fact this is occurring within less than one cloud free-fall time for the simulated cloud lends further credence to SDC335 being at an early period of star formation. (For SDC335 $t_{ff}\sim$ 3.5 $\times 10^5$ yrs.)

Measuring the level of turbulence of material within the central region of SDC335 and the accretion flows onto individual cores would require higher resolution observations of the dense gas (combined with single-dish data to avoid `missing spacing' problems), which are not currently available.

%------------------------------ START Rosen Fig ------------------------------------------
%------------------------------- END Rosen Fig  --------------------------------------------

%-------------------------------------------------------------------------------------------------------------------
%                             Conclusion  
%-------------------------------------------------------------------------------------------------------------------
\section{Discussion and conclusions}
%--- Final remarks,

Using new ATCA SiO and \meth\ observations coupled with archival CO, \thirCO,\ and HNC data from ALMA and four transitions of CO from the APEX telescope, we identify and analyse three molecular outflows within the young high-mass star forming infrared dark cloud SDC335.   These data have yielded the following outcomes: 
\begin{itemize}

\item The three outflows, A, B, and C, are identified,  each associated with one of the three known HC\hii\ regions in SDC335 \citepalias{Avison15} (\mma, \mmb, and MM2,  respectively). The red-blue outflow lobes from A extend in the north-south direction, B extends east-west and C, of which only the blue lobe is detected, extends to the north-west. They have a full width velocity ranges of up to 10\kms\ and temperatures of $\sim$60\,K. The two most massive sources in the cloud, \mma\ and \mmb, are separated by $\sim9000$AU but have driving outflows of projected outflow axes that are approximately perpendicular to one another. The blue lobe of outflow A displays a structure and velocity that is characteristic of a jet.

\item The analysis of the measured outflow momentum flux, $F_{CO}$, as a function of source bolometric luminosity, $L_{bol}$, and in comparison to theoretical evolutionary tracks \citep{Ana13} confirms that the progenitor protostars are massive young stellar objects with two sources residing above the tracks for 50\solmass\ stars. Using samples of $F_{CO}$ and $L_{bol}$ measurements from the literature for low to intermediate mass stars at evolutionary classes 0 and I, we derived best-fit $F_{CO}$-$L_{bol}$ relations for these two classes. Extrapolating these relations upward in $L_{bol}$, we find the outflow momentum flux properties of the SDC335 outflows A and B agree best, with their progenitors being high-mass Class 0 analogues and indicating that SDC335 is at a very early stage of the star-formation process.

\item Inferring the mass accretion rates from the source outflow properties, we find that the total mass accretion is 1.4$(\pm0.1) \times 10^{-3}$\solmass\ yr$^{-1}$ on the protostellar scale. This value is consistent with the calculated mass infall rate on cloud and filamentary scales 2.5$(\pm1.0) \times 10^{-3}$\solmass\ yr$^{-1}$ found by \citet{Peretto13}. This result suggests that at this early stage of evolution nearly all the material accreted onto the clump is funnelled through the cores onto these three massive young sources, limiting the scope for the formation of additional, low mass sources in the region.  If significant numbers of lower mass stars are to form in SDC335, these would then have to form at a later stage in the evolution of the region. 

\item These new data combined with existing knowledge of the bulk inflow and global and filamentary collapse properties of the SDC335 cloud provide compelling evidence of the interaction between the molecular outflows and the material infalling along the filamentary arms. Given the very young (Class 0 analogue) status of the protostars driving the outflows, this makes SDC335 a valuable test bed for study of the disruptive feedback effects of massive protostars on their natal clouds.

\end{itemize}

The observed properties described in this work make the infrared dark cloud SDC335 a key target for the more detailed study of how massive protostars form and effect their natal environments through accretion and protostellar outflows. Such features warrant further study at high spectral and spatial resolutions and sensitivities.

%-------------------------------------------------------------------------------------------------------------------
%                             Acknowledgement   
%-------------------------------------------------------------------------------------------------------------------
\begin{acknowledgements}
The Australia Telescope Compact Array is part of the Australia Telescope which is funded by the Commonwealth of Australia for operation as a National Facility managed by CSIRO. The authors would like to thank all ATNF staff past and present who helped during the ATCA observation used in this paper, particularly those who provided A.A. with curry. The authors would also like to thank the anonymous referee for their input into the paper after initial submission which has helped to improve the work. 

This paper makes use of the following ALMA data: ADS/JAO.ALMA\#2011.0.00474.S and \#2012.0.00781.S. ALMA is a partnership of ESO (representing its member states), NSF (USA) and NINS (Japan), together with NRC (Canada) and NSC and ASIAA (Taiwan), in cooperation with the Republic of Chile. The Joint ALMA Observatory is operated by ESO, AUI/NRAO and NAOJ.  This publication is based on data acquired with the Atacama Pathfinder EXperiment (APEX). APEX is a collaboration between the Max-Planck-Institut fuer Radioastronomie, the European Southern Observatory, and the Onsala Space Observatory. A.A. is funded by the STFC at the UK ARC Node.  G.A.F acknowledges financial support from the State Agency for Research of the Spanish MCIU through the AYA2017-84390-C2-1-R grant (co-funded by FEDER) and through the ``Center of Excellence Severo Ochoa'' award for the Instituto de Astrof\'isica de Andalucia
(SEV-2017-0709). N.P. wishes to acknowledge support under STFC consolidated grants ST/N000706/1 and ST/S00033X/1. A.D.C acknowledges the support from the UK STFC consolidated grant ST/N000706/1. A.L.R acknowledges support from NASA through Einstein Postdoctoral Fellowship grant number PF7- 180166 awarded by the Chandra X-ray Center, which is operated by the Smithsonian Astrophysical Observatory for NASA under contract NAS8-03060. This research made use of APLpy, an open-source plotting package for Python hosted at http://aplpy.github.com. This research made use of Astropy,\footnote{http://www.astropy.org} a community-developed core Python package for Astronomy \citep{astropy:2013, astropy:2018}.
\end{acknowledgements}

%--------------------------------------
% APPENDIX A
%--------------------------------------
\appendix
\section{Spectra of the outflow driving sources}
\label{HCHII_SPEC:app}
\begin{figure}
\centering
\includegraphics[scale=0.55]{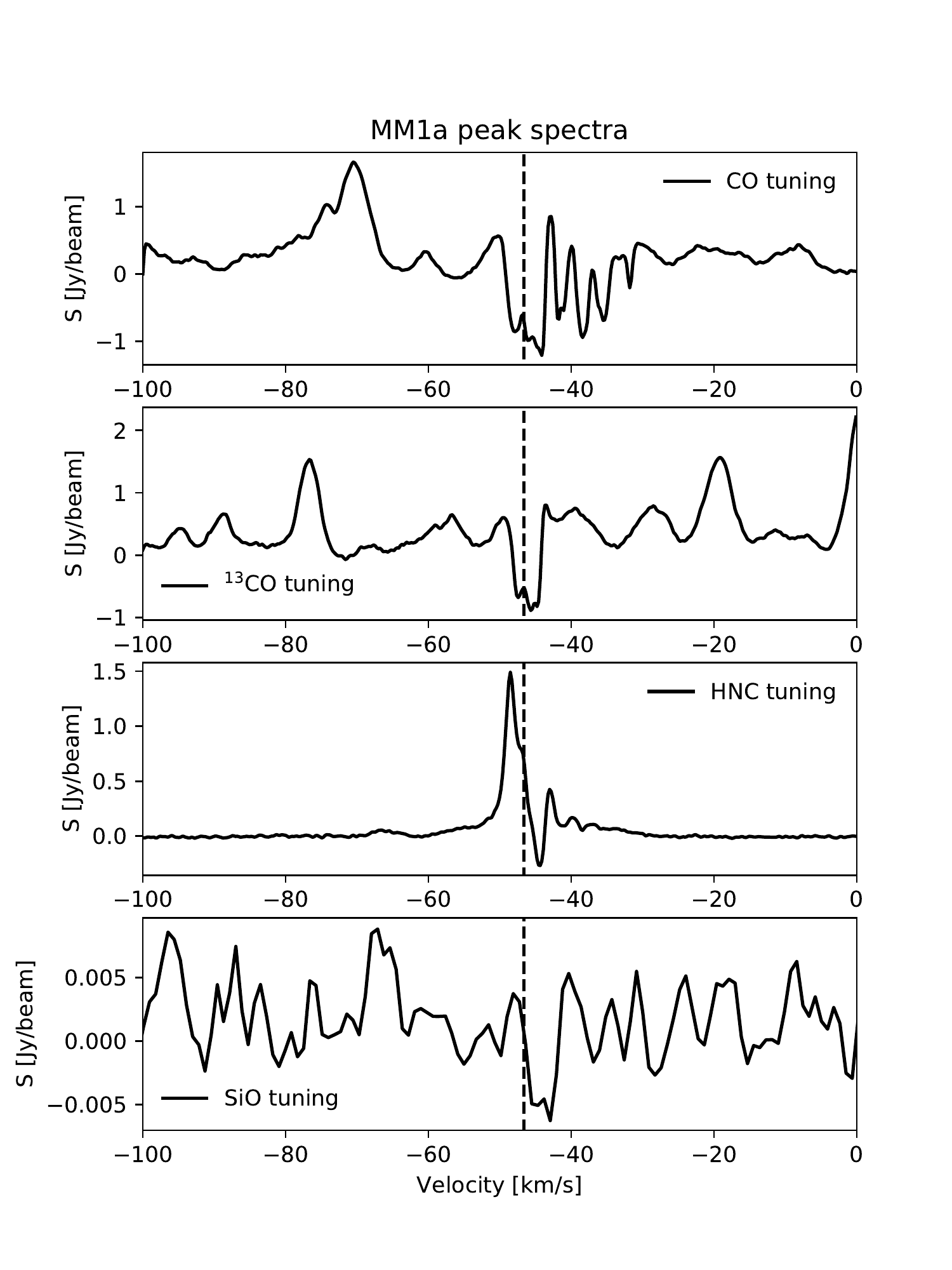}
\caption{Spectra at the position of peak emission at 23GHz \citep{Avison15} from the \mma\ HC\hii\ region at the same CO, \thirCO, HNC, and SiO frequency tunings, as shown in Figures \ref{OUTA_SPEC_blue:fig} to \ref{OUTC_SPEC:fig}.The vertical dashed line gives the $V_{lsr}$ of the HC\hii.}
\label{MM1a_SPEC:fig}%
\end{figure}

\begin{figure}
\centering
\includegraphics[scale=0.55]{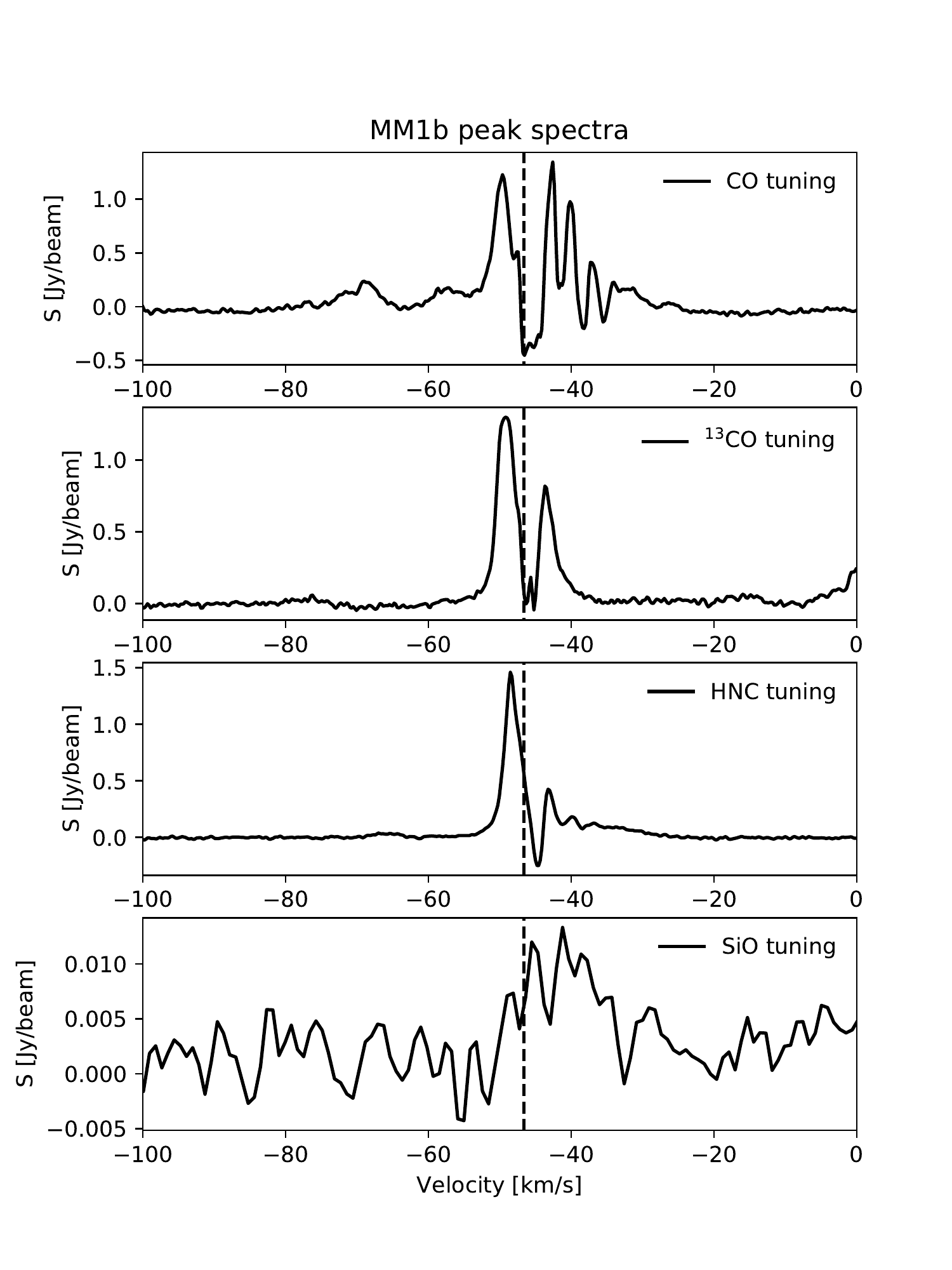}
\caption{As in Figure \ref{MM1a_SPEC:fig}, but at the peak position of \mmb.}
\label{MM1b_SPEC:fig}%
\end{figure}

To compliment the outflow spectra in Figures \ref{OUTA_SPEC_blue:fig} to \ref{OUTC_SPEC:fig}, we present in Figures \ref{MM1a_SPEC:fig} to \ref{MM1b_SPEC:fig} spectra over the same velocity range at the peak position of the three HC\hii\ regions in SDC335. These are the likely driving sources of the outflows, as described in the main text.
 
 \begin{figure}
\centering
\includegraphics[scale=0.55]{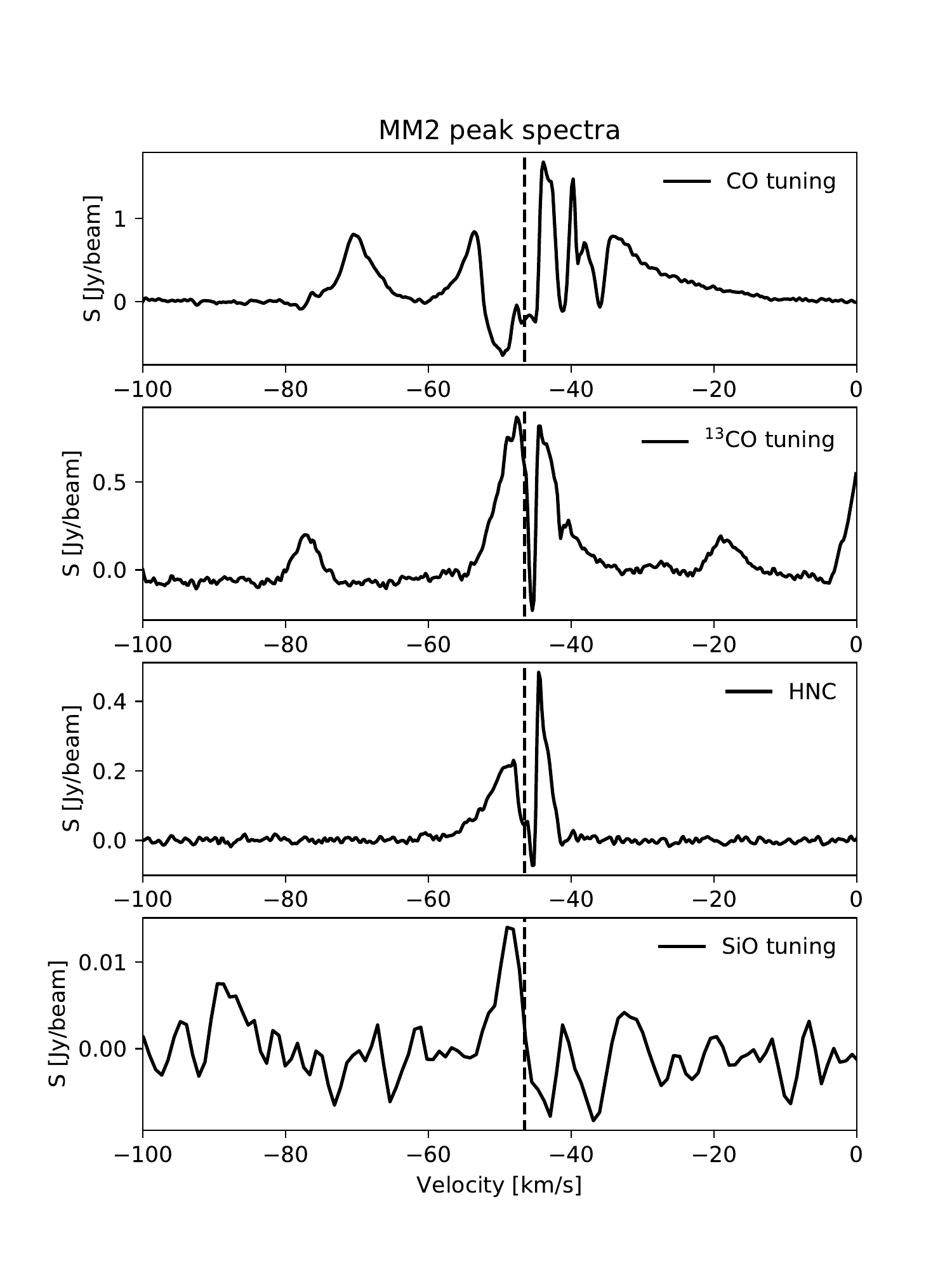}
\caption{As in Figure \ref{MM1a_SPEC:fig}, but at the peak position of MM2.}
\label{MM2_SPEC:fig}%
\end{figure}

%--------------------------------------
% APPENDIX B
%--------------------------------------

\section{Inclination angle considerations based on observed length-to-width ratio}
\label{incAngle:app}
Here, we address the arguments used to limit the range of possible inclination angles derived from the \citep{CabritBertout86} models by introducing a general outflow model beyond the simple bicone.

 \begin{figure*}
\centering
\includegraphics[scale=0.5,  trim = 0 60 40 10, clip]{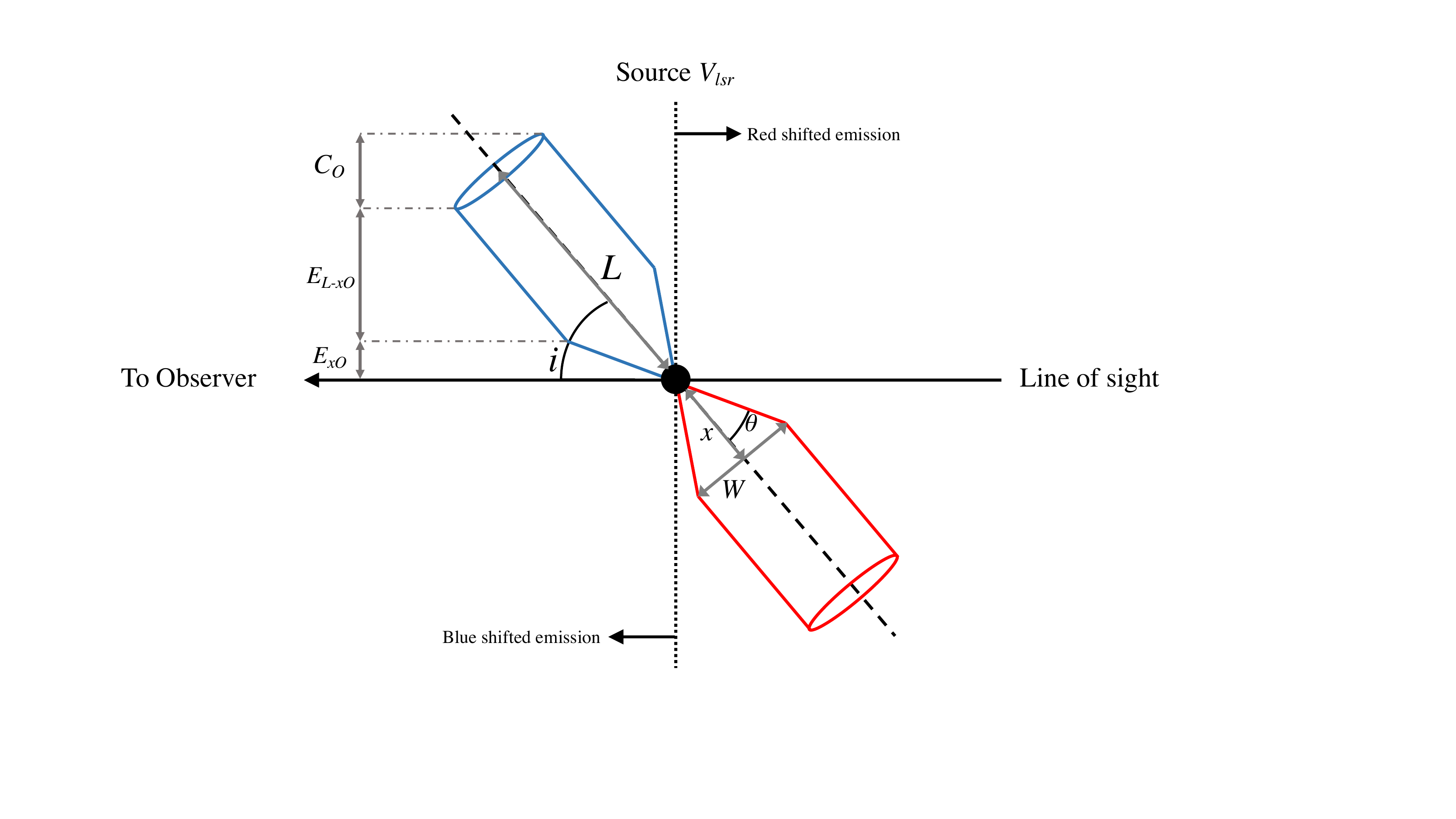}
\caption{Schematic of the parameters for a generalised, non-bicone, outflow model used to limit the possible inclination angles for outflows A and B.}%\LEt{ Please check if this should perhaps be non-bicone.}
\label{NewOutflow:fig}%
\end{figure*}

Following Figure \ref{NewOutflow:fig}, we define the length, $L$, of an outflow as from the driving source to the end of the outflows influence on the surrounding medium in the direction of the outflowing material. Complementary to this, the outflow width, $W$, is the maximum extent of the outflows influence orthogonal to the outflow direction. The angle between the axis of symmetry of the outflow (along $L$) and the widest point, called the opening angle, we denote here as $\theta$. The distance along $L$ to the point of maximum width we name $x$. The length of $x$ depends on the morphology of the outflow. In the case of the \citep{CabritBertout86} bicone model $L = x$. The true length-to-width, $LWR$, of the outflow is then defined as $LWR = L/W = \frac{L}{2x\tan{\theta}}$. 

Also presented in Figure \ref{NewOutflow:fig} are the observed properties of the outflow. The observed length $L_O$ of the outflow at an inclination angle, $i$, with respect to the observer is the sum of the contribution of the outflow wall or edge, $E_O$, and emission from the outflow cavity/dome end, $C_O$. So, $L_O = E_O + C_O$. 

In turn $E_O$ comprises the sum of the outflow wall from it origin to length $x$, $E_{xO}$ and from the wall from $x$ to the end of the outflow at $L$, $E_{L-xO}$. These two contributions have the forms:

\begin{equation}
E_{xO} = E_x \sin(i-\theta) = \frac{W}{2\sin(\theta)} \sin(i-\theta)
,\end{equation}

\begin{equation}
E_{L - xO} = (L-x) \sin i
,\end{equation}

\noindent leading to

\begin{equation}
E_{O} = \frac{W}{2\sin(\theta)} \sin(i-\theta) + (L-x) \sin i 
\end{equation}

\noindent and the contribution of the cavity along the line of sight is given by,
\begin{equation}
 C_O = W\cos i 
,\end{equation} 

\noindent such that in its full form the observed length is given by
\begin{equation}
L_O =   \frac{W}{2\sin(\theta)} \sin(i-\theta) + (L-x) \sin i +W \cos i \rm{~}.
\label{obsLen:eqn}
\end{equation} 
 
We note that inclination projection effects on $W$ are minimal, for example, given a source at distance of 3kpc the observed width only differs from the true $W$ due to its inclination at the 0.02\% level for $L$ = 0.5pc (larger than any measured or corrected length in our observed outflows) when inclined along the line of sight. Indeed, an outflow of $L$ = 30pc inclined along the line of sight will only have a project effect difference on $W$ at the 1\% level. As such, we neglect this consideration. This gives the observed length-to-width as $LWR_O = \frac{L_O}{W}$. There is a final observable counterpart to the true physical value which is the observed length from the driving source to the widest point, $x_O$ which is given by  $x \sin i$.

\subsection{Bicone case, $L = x$}
In the case that the outflow has a biconical morphology, as is the case assumed to derive our initial inclination angle ranges, then $L=x$ and the second term in Equation \ref{obsLen:eqn} becomes zero. This leads the observed length-to-width to become:

\begin{equation}
LWR_O =   \frac{1}{2\sin(\theta)} \sin(i-\theta) + \cos i 
\label{LWRbiconen:eqn}
,\end{equation} 

\noindent from which using the measured $LWR_O$ and $\theta$ it is possible to constrain the value of inclination angle, $i$. For outflow A with a $\theta$ = 11\degs\ and $LWR_O$ = 3.5, Equation \ref{LWRbiconen:eqn} asymptotes at 79\degs\ with a $LWR_O$ = 2.6. Similarly for outflow B with a $\theta$ = 14\degs\ and $LWR_O$ = 3.0 asymptotes at 76\degs\ with a $LWR_O$ = 2.1. In both cases, these asymptotes are clearly at 90\degs\ - $\theta$. Based on this, it becomes clear that whilst the biconical morphology does not apply particularly well to outflows A and B, the observed length-to-width ratios are providing evidence that (in the absence of external influence on the morphology) these outflows tend toward higher inclination angles.

\subsection{General case, $L \neq x$}
\label{generalCase:sec}
For the general case, where $L \neq x$, the inspection of the inclination angle now depends on an additional piece of information, the observed distance from the outflow origin to the point that the outflow is widest, $x_O$. With this the observed $LWR_O$ becomes,

\begin{equation}
LWR_O =   \frac{1}{2\sin(\theta)} \sin(i-\theta) + \frac{(L-\frac{\sin i}{x_0}) \sin i}{W} + \cos i \rm{~}.
\label{LWRgen:eqn}
\end{equation}

Given our observed $L_O$, $W$, $\theta$ and $x_O,$ we can solve numerically the range of $i$ and $x$ values that will return a $LWR$ matching our observed values. We do this for a model outflow of $L = 1$ with $x$ and $W$ as fractions of this. For this work, we consider values of $i$ and $x$ which return both an $x_O$ and $LWR_O$ within $\pm 10\%$ of our observed values. 

Under these conditions we find that we can limit the range of $i$ values to between 53 and 76\degs\ for outflow A and 59 and 89\degs\ for outflow B. The latter values can again be limited at the higher end by kinematics, as above 79\degs\ we would observe both red- and blue-shifted emission at either side of the driving source for outflow B, giving us a final limit for B of between 59 and 79\degs. These are significantly smaller ranges than using the \citet{CabritBertout86} models and again tend to higher inclination angles. 

\subsection{Implications of using the bicone model}
In Section \ref{angConsider:sec}, the morphology described in Section \ref{generalCase:sec} is used in place of the typical biconical outflow morphology to narrow the ranges of potential inclination angles for outflows A and B observed in SDC335. Table \ref{CorrectionFactors:tab} provides the factors required to correct the observed values for the effects of inclination. These correction factors are given for both the angle ranges generated using our preferred morphology and the bicone case.

Owing to the difference in angles covered the correction factors differ between the two morphologies by factors of 0.65 and 2.15 for outflow A and 0.6 and 2.65 for outflow B, respectively.  One item that is of significance with regard to the findings of this paper is the correction factor applied to the momentum flux, $F$. Using the bicone morphology, the momentum flux values are 2.15$\times$ and 2.65$\times$ smaller for outflows A and B, respectively. Following this reduction through our analysis yields the following changes:
\begin{itemize}

\item The $F_{CO}$ values for \mma\ and \mmb\ become 1161.8$\times 10^{-5}$\solmass\ \kms\ yr$^{-1}$ and 341.3$\times 10^{-5}$\solmass\ \kms\ yr$^{-1}$, respectively (for the sum of the red and blue lobes), leading to each source being lower on the y-axis of Figure \ref{Fout_Lbol:fig}. \mma\ remains consistent with the Class 0 line of best fit whereas \mmb\ moves to the upper end of the 1$-\sigma$ error margin for the Class I line of best fit, suggesting it is potentially more evolved than \mma\, though remains consistent with the Class 0 line within errors. Both sources remain consistent with the 50\solmass\ evolutionary track plotted in \ref{Fout_Lbol:fig}.

\item The $\dot M_{acc}$ derived from our $F_{CO}$ values are now in the range 6.8 - 39.6 $\times 10^{-5}$\solmass\ yr$^{-1}$ , at $T=20$K, $\tau = 3.5$, and 4.3 - 9.3 $\times 10^{-5}$\solmass\ yr$^{-1}$, for $T=53-62$K with $\tau = 8.2$ for outflows A and B. From these, the total derived mass accretion, $\dot M_{tot. acc}$, within SDC335 (as defined in \S \ref{infall:sec}) becomes $6.04 (\pm 0.04) \times 10^{-4}$\solmass\ yr$^{-1}$. This means the comparison to the total infall rate is somewhat weaker, at 24\% of the \citet{Peretto13} values, rather than 55\% and comparable within the errors.

\item The change in $\dot M_{acc}$ leads to a lessening of the discrepancy discussed in \S \ref{tooLow:sec} between the bolometric luminosity ,$L_{bol}$, and $L_{tot,ZAMS}$. For \mmb\ with $L_{tot,ZAMS},$ becomes a factor 1.5-2.4 times lower than $L_{bol}$, meaning these values are consistent (as is the case with source MM2).  For \mma\ range of  $\dot M_{acc}$ values, an allowance remains for a discrepancy of greater than 2.5$\times$ between the two luminosities at the higher end ($>30\times10^{-5}$\solmass\ yr$^{-1}$) of the derived range.
\end{itemize}
%-------------------------------------------------------------------------------------------------------------------
%                             Bibliography   
%-------------------------------------------------------------------------------------------------------------------
\bibliographystyle{aa}
\bibliography{SDC335_OUT_toArXiv.bib}

\end{document}